\documentclass[fleqn,usenatbib]{mnras}
\usepackage{newtxtext,newtxmath}
\usepackage[T1]{fontenc}
\DeclareRobustCommand{\VAN}[3]{#2}
\let\VANthebibliography\thebibliography
\def\thebibliography{\DeclareRobustCommand{\VAN}[3]{##3}\VANthebibliography}

\usepackage{graphicx}	
\usepackage{amsmath}	
\usepackage{multirow}
\usepackage{pdflscape}
\usepackage{threeparttable}
\newcommand\MF{CH$_{3}$OCHO}
\newcommand\DE{CH$_{3}$OCH$_{3}$}
\newcommand\K{H$_{2}$CCO}

\title[Correlations]{Correlations of Methyl Formate (\MF), Dimethyl Ether (\DE) and Ketene (\K) in High-mass Star-forming Regions}

\author[Li et al.]{
Chuanshou Li,$^{1}$\thanks{E-mail: lichuanshou2021@163.com}
Sheng-Li Qin,$^{1}$
Tie Liu,$^{2}$
Sheng-Yuan Liu,$^{3}$
Mengyao Tang,$^{4}$
Hong-Li Liu,$^{1}$
Li Chen,$^{1}$
\newauthor
Xiaohu Li,$^{5}$
Fengwei Xu$^{6,7}$
Tianwei Zhang,$^{8,9}$
Meizhu Liu,$^{1}$
Hongqiong Shi,$^{1}$
and Yuefang Wu$^{7}$
\\
$^{1}$School of Physics and Astronomy, Yunnan University, Kunming 650091, People's Republic of China\\
$^{2}$Shanghai Astronomical Observatory, Chinese Academy of Sciences, 80 Nandan Road, Shanghai 200030, People's Republic of China\\
$^{3}$Academia Sinica Institute of Astronomy and Astrophysics, 11F AS/NTU Astronomy-Mathematics Building, No. 1, Section 4, Roosevelt Road, Taipei 10617, Taiwan\\
$^{4}$Institute of Astrophysics, School of Physics and Electronic Science, Chuxiong Normal University, Chuxiong 675000, People’s Republic of China\\
$^{5}$Xinjiang Astronomical Observatory, Chinese Academy of Sciences, Urumqi 831399, China\\
$^{6}$Kavli Institute for Astronomy and Astrophysics, Peking University, Beĳing, 100871, People's Republic of China\\
$^{7}$Department of Astronomy, School of Physics, Peking University, Beĳing, 100871, People's Republic of China\\
$^{8}$I. Physikalisches Institut, Universit{\"a}t zu K{\"o}ln, Z{\"u}lpicher Stra{\ss}e 77, 50937 K{\"o}ln, Germany\\
$^{9}$Research Center for Intelligent Computing Platforms, Zhejiang Laboratory, Hangzhou 311100, P.R.China
}

\date{Accepted XXX. Received YYY; in original form ZZZ}

\pubyear{2024}

\begin{document}
\label{firstpage}
\pagerange{\pageref{firstpage}--\pageref{lastpage}}
\maketitle

\begin{abstract}
We present high-spatial-resolution (0.7 to 1.0 arcsec) submillimeter observations of continuum and molecular lines of \MF, \DE, and \K\ toward 11 high-mass star-forming regions using the Atacama Large Millimetre/submillimetre Array (ALMA). A total of 19 separate cores from 9 high-mass star-forming regions are found to be line-rich, including high-, intermediate-, and low-mass line-rich cores. The three molecules are detected in these line-rich cores. We map the emission of \MF, \DE, and \K\ in 9 high-mass star-forming regions. The spatial distribution of the three molecules is very similar and concentrated in the areas of intense continuum emission. We also calculate the rotation temperatures, column densities, and abundances of \MF, \DE, and \K\ under the local thermodynamic equilibrium (LTE) assumption. The abundances relative to H$_{2}$ and CH$_{3}$OH, and line widths of the three molecules are significantly correlated. The abundances relative to H$_{2}$, temperatures and line widths of the three molecules tend to be higher in cores with higher mass and outflows detected. The possible chemical links of the three molecules are discussed.
\end{abstract}


\section{Introduction} \label{sec:Introduction}
The astrochemical networks of many species have gradually been revealed. These species range from simple neutral molecules, molecular radicals, and ions to complex organic molecules (COMs). COMs are defined as C-bearing molecules with at least six atoms \citep{2009ARA&A..47..427H}. At present, two major pathways for producing organic molecules have been proposed: (i) grain-surface chemical reactions \citep{1992ApJS...82..167H, 2000MNRAS.319..837R, 2008ApJ...682..283G,  2015MNRAS.447.4004R}; (ii) gas-phase chemical reactions \citep{1984inch.book.....D, 2013ApJ...769...34V, 2015MNRAS.449L..16B}. Comparing the abundance and spatial distribution correlations of different species is an important mean to test chemical models and determine their formation paths. For example, interferometric observations showed a difference in the spatial distribution of O- and N-bearing molecules \citep{2008ApJ...672..962F, 2019A&A...632A..57C, 2015ApJ...803...39Q, 2022MNRAS.511.3463Q}, with N-bearing molecules tracing higher temperature gas than O-bearing molecules \citep{2010ApJ...711..399Q, 2015ApJ...806..239C, 2020ApJ...897L..38V}. In particular, methyl formate (\MF) and dimethyl ether (\DE) were found to have a potential similarity \citep{2014ApJ...791...29J, 2020A&A...641A..54C, 2022MNRAS.512.4419P, 2023A&A...678A.137C}. Nevertheless, the spatial similarity between \MF\ and \DE\ has not been systematically confirmed by interferometric observations with large samples.

Ketene (\K) has been proposed as an important precursor for the formation of COMs, such as acetic acid (CH$_{3}$COOH), acetamide (CH$_{3}$CONH$_{2}$), pyruvonitrile (CH$_{3}$COCN) and methyl acetate (CH$_{3}$COOCH$_{3}$) \citep{2013ApJ...773..109H}. However, there are still many uncertainties about how this molecule is produced \citep{2005IAUS..231..237C, 2007Ap&SS.310..181R, 2013ApJ...769...34V, 2014ApJ...789...36M, 2017ApJ...847...89K, 2022ApJ...924..110F}. \MF, \DE, and \K\ are commonly detected in the same sources of hot molecular cores (HMCs) \citep{2000ApJS..128..213N, 2013A&A...559A..47B}, hot corinos \citep{2017ApJ...841..120B, 2018A&A...620A.170J, 2019ESC.....3.1564B, 2019A&A...625A.147A} or cold regions \citep{2012A&A...541L..12B, 2012ApJ...759L..43C, 2014ApJ...791...29J}. Previous models have also shown that there may be chemical associations among the three molecules \citep{2006A&A...457..927G, 2008ApJ...682..283G, 2015MNRAS.449L..16B, 2001AcSpA..57..685C, 2005IAUS..231..237C}. Whereas the relationship of \K\ with the other two molecules has been barely explored from the observational side.

In this work, the correlations among \MF, \DE, and \K\ are investigated toward 11 high-mass star-forming regions using Atacama Large Millimeter/submillimeter Array (ALMA) data. These data were observed as a pilot project for the ALMA Three-millimetre Observations of Massive Star-forming regions (ATOMS) survey \citep{2020MNRAS.496.2790L}. These 11 targets have large masses and luminosity \citep{2024ApJS..270....9X}, and show infall motion traced by the ``blue profiles'' observed by the HCN (4-3) lines \citep{2016ApJ...829...59L}. The paper is organized as follows. The observations and data reduction are described in Section \ref{sec:Observations}. The observational results including continuum, line identifications, parameter calculation, and molecular emission maps are given in Section \ref{sec:Results}. We discuss the correlations of the three molecules and their implications for chemistry in Section \ref{sec:Discussion}. The main results and conclusions are summarized in Section \ref{sec:Conclusions}.

\section{Observations} \label{sec:Observations}
The basic observational parameters and data reduction are described in \cite{2024ApJ...962...13C}. Observations of 11 high-mass star-forming regions were performed from 18 to 20 May 2018 under the ALMA Cycle 5 project at Band 7 (ID: 2017.1.00545.S, PI: Tie Liu). A total of 43 antennas with a diameter of 12m were employed for observations (C43-1 configuration). The Band 7 observations offer 4 spectral windows (SPWs) centering at the frequency of 343.2 (SPW 31), 345.1 (SPW 29), 345.4 (SPW 25), and 356.7 GHz (SPW 27), respectively. Both SPWs 25 and 27 have a bandwidth of 469 MHz and a spectral resolution of 0.24 km s$^{-1}$, and are used to observe HCN (4-3) and HCO$^{+}$ (4-3) lines, respectively. In this paper, we present the spectral line data of SPW 29 and SPW 31, both having a bandwidth of 1.88 GHz with a velocity resolution of 0.98 km s$^{-1}$. We used the TCLEAN task in Common Astronomy Software Applications (CASA; \citealt{2007ASPC..376..127M}) to generate images of continuum and spectral cubes. Before the imaging processes, the bandpass, amplitude, and phase calibration were performed by running ``ScriptForPI.py'' offered by the ALMA team. For I14498, the phase calibrator is J1524-5903, and the flux and bandpass calibrator is J1427-4206. For I17220, the phase calibrator is J1733-3722, and the flux and bandpass calibrator is J1924-2914. As for the other 7 sources, the phase calibrator is J1650-5044, and the flux and bandpass calibrators are J1924-2914, J1427-4206, and J1517-2422. The continuum image was constructed from all line-free channels of four spectral windows. Self-calibrations were performed to improve the qualities of continuum images. Totally 3 rounds of phase self-calibrations and 1 round of amplitude self-calibration had been conducted to all continuum data of 11 sources. During the self-calibration, we used the ``hogbom''  deconvolution algorythm with a weighting parameter of  ``briggs''. The robust parameter of ``briggs'' weighting can be set from $-$2.0 to 2.0 to smoothly customize trade-offs between resolution and imaging sensitivity. The ``robuts=0.5'' was set to balance the sensitivity and resolution of our images. Following self-calibration, the Root-Mean-Square (rms) noise of the source I15520 was reduced by approximately 19\%, the source I14498, I16351 and I17220 by about 3\%, and the source I15596, I16060, I16076, I16272 and I17204 were barely improved. The primary beam correction had also been performed. Finally, the solutions from the self-calibration of continuum images were applied to line cubes. The synthesized beam sizes of continuum images and line cubes range from 0.7 to 1.0 arcsec. The average sensitivity is better than 8.0 mJy beam$^{-1}$ per channel for line cubes, and better than 2.5 mJy beam$^{-1}$ for continuum images.

\section{Results} \label{sec:Results}
\subsection{Continuum Emission} \label{subsec:ce}
\begin{figure*}
	\includegraphics[width=\linewidth]{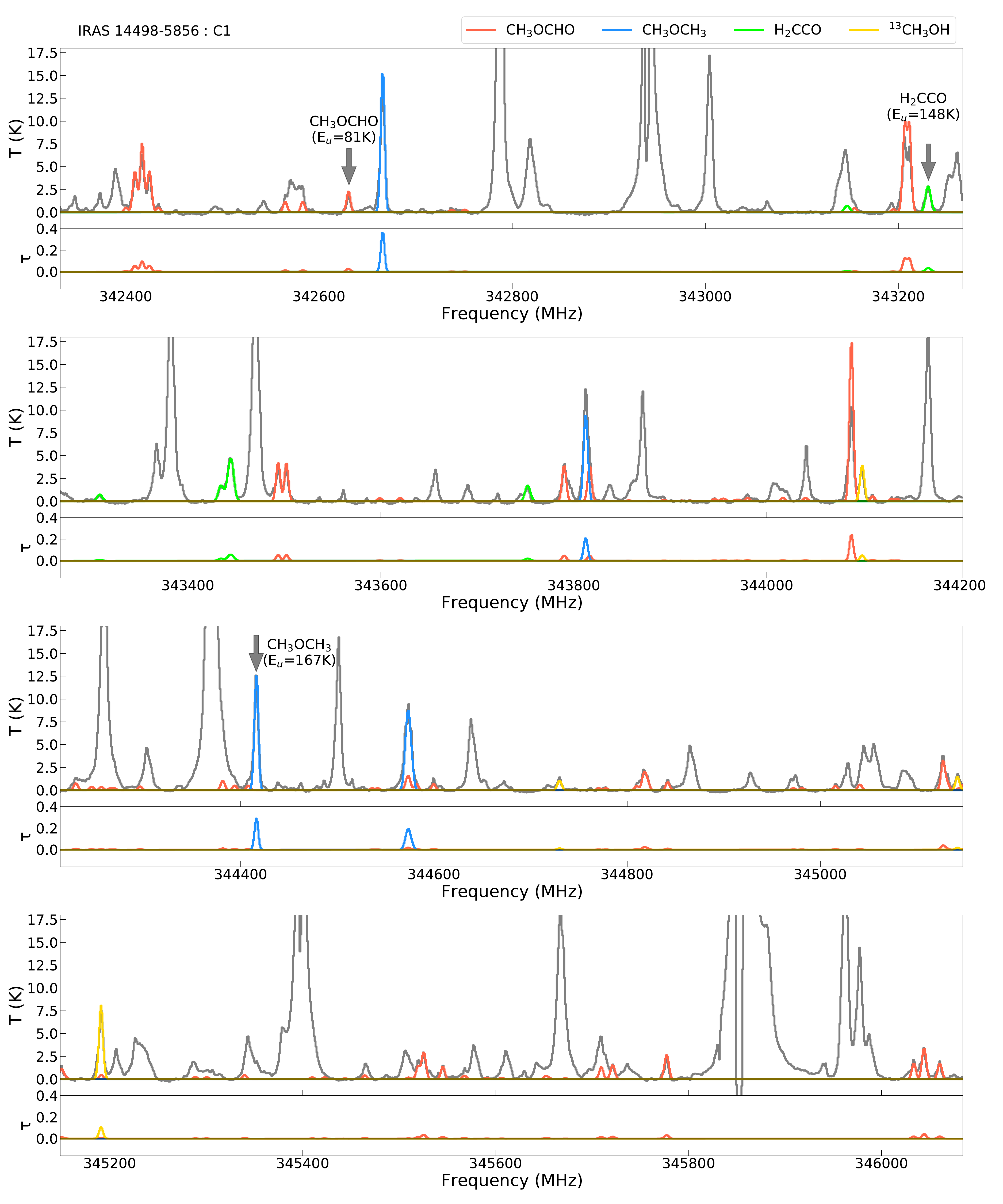}
	\caption{Sample spectra and optical depths of the three molecules for I14498 C1. The observed spectra are shown in gray curves and the XCLASS modeled spectra are shown in color curves. The small panel below each spectrum shows the optical depths of the XCLASS modeled spectra. The gray arrows represent the transitions used as integrated intensity maps in Figure \ref{fig2}. The spectra and optical depths of other line-rich cores are shown in Figure \ref{fig8}.} \label{fig1}
\end{figure*}

We perform multi-component two-dimensional Gaussian fits to 870 $\mu$m continuum emission using the CASA-$imfit$ function. A total of 145 dense cores were resolved in 11 high-mass star-forming regions \citep{2024ApJ...962...13C}. We have checked the spectra towards the 145 cores one by one, and found that 19 separate cores in 9 regions have multiple line emissions of \MF\ or \DE\ or \K. From Figure \ref{fig1} and \ref{fig8}, these 19 cores show rich line emission. The ALMA 870 $\mu$m continuum emission from 9 high-mass star-forming regions is shown in Figure \ref{fig2}. The positions of 19 line-rich cores are labeled. For the other two of the 11 high-mass star-forming regions (IRAS 14382-6017 and IRAS 17204-3636), the line transitions of the three molecules were not detected. The absence of HMCs from these two sources was affirmative by cross-matching with the ALMA Band 3 dataset \citep{2022MNRAS.511.3463Q}.

For the estimation of the core masses and the molecular hydrogen column densities, the commonly used method assumes that the dust emission is optically thin and in local thermodynamic equilibrium (LTE). Under these assumptions, the core masses and source-averaged H$_{2}$ column densities can be estimated by the expressions \citep{2008A&A...487..993K}:
\begin{equation} \label{eq1}
	\rm M_{core} = \frac{S_{\nu} \eta D^{2}}{\kappa_{\nu} B_{\nu}(T)},
\end{equation}
\begin{equation} \label{eq2}
	\rm N (H_{2}) = \frac{S_{\nu} \eta}{\mu m_{H} \Omega \kappa_{\nu} B_{\nu}(T)},
\end{equation}
where S$_{\nu}$ is integrated flux density, $\eta$ = 100 is gas-to-dust mass ratio \citep{1991ApJ...380..429L,1992ApJS...82..167H}, D is the distance to the core \citep{2020MNRAS.496.2790L} and the uncertainty takes 10\% \citep{2020ApJ...890...44B, 2022MNRAS.511..501L} when calculating the uncertainty of the core mass, $\kappa$$_{\nu}$ is the dust mass absorption coefficient, B$_{\nu}$(T) is the Planck function at dust temperature T, $\mu$ = 2.8 is the mean molecular weight of the gas \citep{2008A&A...487..993K}, m$_{\rm H}$ is the mass of the hydrogen atom, and $\Omega$ is the solid angle corresponding to the deconvolved size of the core. In our case, $\kappa$$_{870\mu m}$ takes a value of 1.89 cm$^{2}$ g$^{-1}$, which is interpolated from the table given in \cite{1994A&A...291..943O}, assuming grains with thin ice mantles and a gas density of 10$^{6}$ cm$^{-3}$. The dust temperatures of dense cores are assumed to be the rotation temperatures of \MF, which is considered to be dust temperature probe \citep{2011A&A...532A..32F}. The parameters of 19 line-rich cores are listed in Table \ref{tab1}. In Table \ref{tab1}, we also list whether these cores are associated with outflows. The presence of outflows is identified by searching for red-blue lobes around continuum sources using the CO (3-2), HCN (4-3), and SiO (2-1) lines \citep{2020ApJ...890...44B}.

\begin{figure*}
	\includegraphics[width=\linewidth]{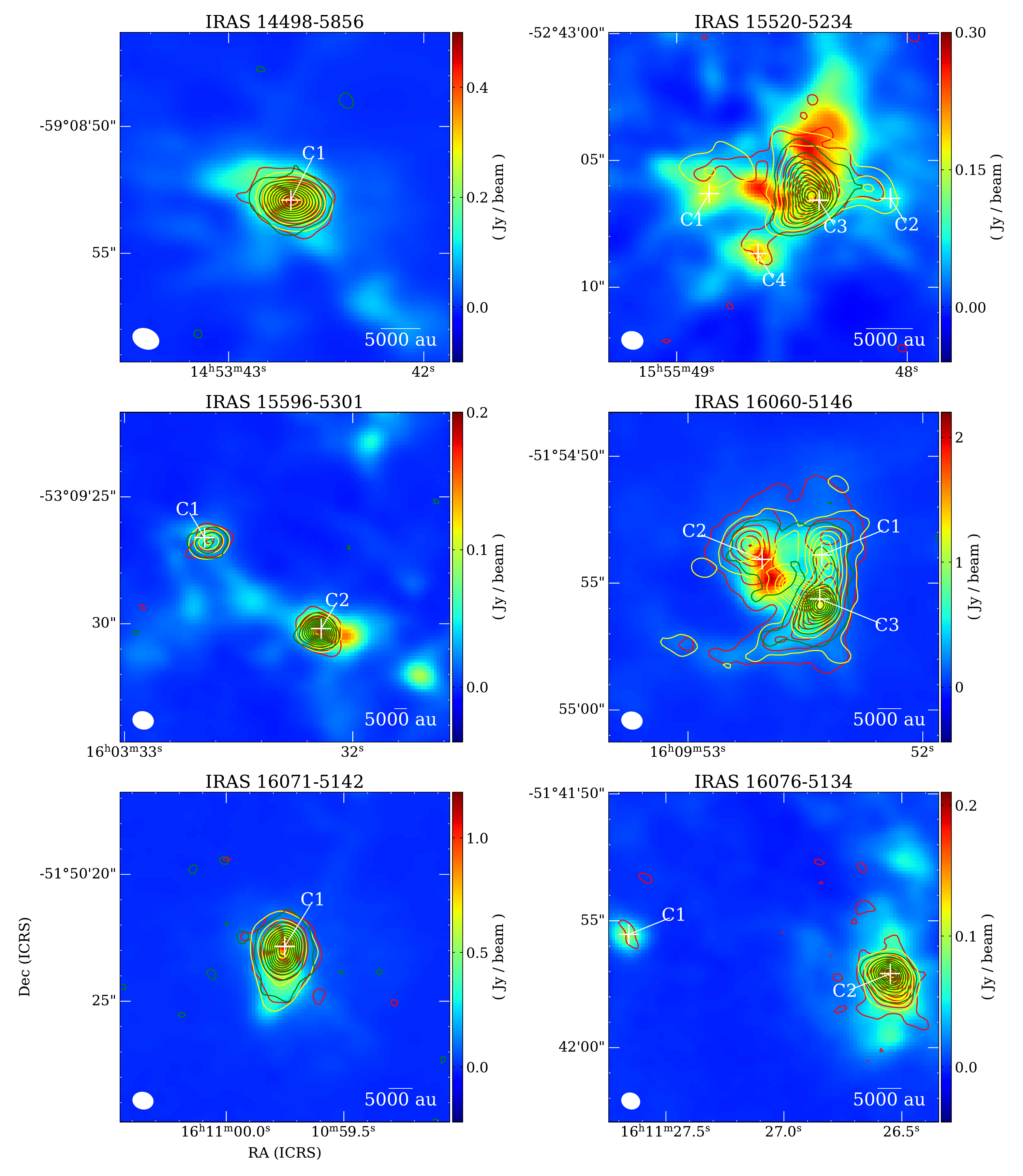}
	\caption{Continuum emission at 870 $\mu$m overlaid with integrated intensities of \MF, \DE, and \K\ in 9 high-mass star-forming regions. The dense cores are labeled by declination (Dec) order. The green, yellow, and red contours represent \MF\ at 342572 MHz (E$_{\rm u}$ = 81 K), \DE\ at 344358 MHz (E$_{\rm u}$ = 167 K), and \K\ at 343173 MHz (E$_{\rm u}$ = 148 K), respectively. The contour levels are stepped by 10\% of the peak values, with the outermost contour levels as follows: I14498: 5\%, I15520: 3\%, I15596: 8\%, I16060: 2\%, I16071: 2\%, I16076: 6\%, I16272: 2\%, I16351: 2\%, I17220: 2\% of the peak values. The synthesized beam sizes are shown on the lower left, the scalebars are shown on the lower right, and the continuum emission colorbars are shown on the right.} \label{fig2}
\end{figure*}

\addtocounter{figure}{-1}
\begin{figure*}
	\includegraphics[width=\linewidth]{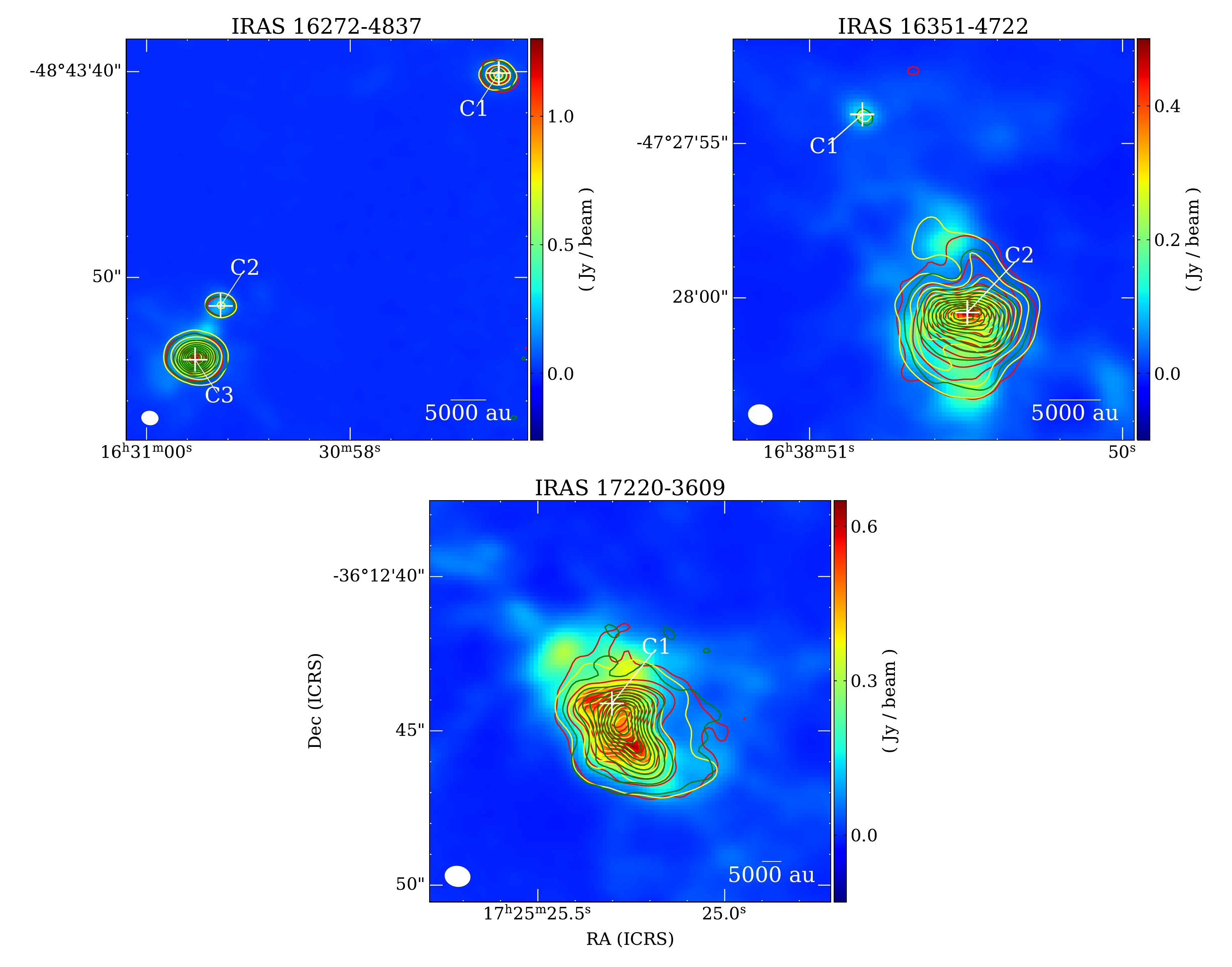}
	\caption{\it -- continued}
\end{figure*}

The source-averaged optical depths of the continuum can be calculated by the following formula \citep{2010ApJ...723.1665F, 2021A&A...648A..66G}:
\begin{equation} \label{eq3}
	\rm \tau_{\nu} = -ln \left( 1 - \frac{S_{\nu}}{\Omega B_{\nu}(T)} \right).
\end{equation}
The derived $\tau$$_{870 \mu m}$ ranges from 6.2 × 10$^{-3}$ to 1.4 × 10$^{-1}$ for the 19 dense cores, so the optically thin assumption is reasonable. I15520, I16060, I16071, I16076, I16351, and I17220 were found to be associated with the ultra-compact (UC) H{\sc ii} regions traced by the H40$\alpha$ lines  \citep{2022MNRAS.511.3463Q, 2022MNRAS.511..501L}. Free-free emissions from UC H{\sc ii} regions in the submillimeter band are weak, which have less contribution to the continuum flux.

\subsection{Line Identifications} \label{subsec:li}
We extracted the spectra from the continuum peak positions of these dense cores. Then the spectral line transitions are identified using the eXtended CASA Line Analysis Software Suite (XCLASS\footnote{\url{https://xclass.astro.uni-koeln.de}}; \citealt{2017A&A...598A...7M}). XCLASS searches for molecular line parameters in the Jet Propulsion Laboratory (JPL\footnote{\url{http:///spec.jpl.nasa.gov}}; \citealt{1998JQSRT..60..883P}) and the Cologne Database for Molecular Spectroscopy (CDMS\footnote{\url{http://cdms.de}}; \citealt{2001A&A...370L..49M, 2005JMoSt.742..215M}). The value of the partition functions in the XCLASS have 110 different temperature intervals between 1.072 and 1000 K. Assuming that the molecular gas satisfies the LTE condition, the XCLASS solves the radiative transfer equation and produces synthetic spectra for specific molecular transitions by taking into account beam dilution, dust attenuation, line opacity, and line blending. In the XCLASS modeling, the input parameters are the size ($\theta$), rotational temperature (T), column density (N), line width ($\Delta$V), and velocity offset (V$_{\rm off}$) of each molecule. In our study, we took the deconvolved sizes (see Table \ref{tab1}) of the continuum sources as molecular component sizes. To better fit the rotation temperature, column density, and line width parameters, we further employed Modeling and Analysis Generic Interface for eXternal numerical codes (MAGIX; \citealt{2013A&A...549A..21M}). The MAGIX optimizes these molecular component parameters within the given range and provides the corresponding error estimates. The parameter ranges are set based on the initial guesses provided by XCLASS fitting. Note that in our case the optical depths of the continuum cores are less than 1.4 × 10$^{-1}$ and then the dust attenuation effect can be ignored.

Transitions are considered as detection if the line intensities exceed 3$\sigma$ noise level. \MF, \DE, \K, and CH$_{3}$OH lines are detected in all 19 line-rich cores, and $^{13}$CH$_{3}$OH lines are detected in 16 line-rich cores. Figure \ref{fig1} shows the sample spectra and optical depths of molecular transitions toward I14498 C1. The spectra and optical depths of molecular transitions toward the other 18 cores are presented in Figure \ref{fig8} of Appendix \ref{sec:solc}. The overall results of the detected spectral lines are summarized as follows:

\begin{landscape}
	\begin{table}
		\centering
		\caption{Physical Parameters of the Continuum Sources.}
		\label{tab1}
		\begin{tabular}{lccccccccccccc}
			\hline\hline
			\multirow{2}{*}{Region} & \multirow{2}{*}{Core} & D & R.A. & DEC. & $\theta$$_{\rm dec}$ & R$_{\rm core}$ & I$_{\rm peak}$ & S$_{\nu}$ & T & M$_{\rm core}$ & N (H$_2$) & \multirow{2}{*}{Outflows} & \multirow{2}{*}{Mass Classification}\\
			&& (kpc) & (h : m : s) & ($\circ$ : $\prime$ : {$\prime$$\prime$}) & maj($\prime$$\prime$) × min($\prime$$\prime$) & (au) & (mJy beam$^{-1}$) & (mJy) & (K) & (M$_{\bigodot}$) & (cm$^{-2}$) &&\\
			\hline
			\multirow{1}{*}{I14498-5856} & C1 &	\multirow{1}{*}{3.2}
			&	14:53:42.7	&	$-$59:08:52.8	&	3.2×1.9	&	7890	&	340±22	&	2830±200	&	102±6	&	21.4±2.9	&	(2.1±0.2)e+23	&	1	&	H	\\
			\hline	
			\multirow{4}{*}{I15520-5234} & C1 &	\multirow{4}{*}{2.7}
			&	15:55:48.9	&	$-$52:43:06.1	&	2.9×2.0	&	6502	&	149±11	&	1600±120	&	80±10	&	11.2±2	&	(1.6±0.2)e+23	&	0	&	H	\\
			&	C2	&&	15:55:48.1	&	$-$52:43:06.5	&	1.9×1.4	&	4404	&	80±3	&	440±18	&	70±36	&	3.6±1.9	&	(1.1±0.6)e+23	&	0	&	I	\\
			&	C3	&&	15:55:48.4	&	$-$52:43:06.5	&	2.0×1.4	&	4518	&	200±5	&	1128±35	&	102±4	&	6.1±0.7	&	(1.8±0.1)e+23	&	1	&	I	\\
			&	C4	&&	15:55:48.6	&	$-$52:43:08.5	&	3.1×1.9	&	6553	&	142±17	&	1510±190	&	105±37	&	7.9±3	&	(1.1±0.4)e+23	&	1	&	I	\\
			\hline	
			\multirow{2}{*}{I15596-5301} & C1 &	\multirow{2}{*}{10.1}
			&	16:03:32.6	&	$-$53:09:26.6	&	1.3×0.8	&	10300	&	55±4	&	160±16	&	100±23	&	12.3±3.3	&	(7.0±1.8)e+22	&	1	&	H	\\
			&	C2	&&	16:03:32.1	&	$-$53:09:30.3	&	2.2×0.9	&	14212	&	166±9	&	799±54	&	110±39	&	55.4±20.7	&	(1.7±0.6)e+23	&	0	&	H	\\
			\hline															
			\multirow{3}{*}{I16060-5146} & C1 &	\multirow{3}{*}{5.3}
			&	16:09:52.4	&	$-$51:54:53.8	&	2.2×1.2	&	8611	&	843±48	&	5000±330	&	112±7	&	93.6±12.7	&	(7.6±0.7)e+23	&	0	&	H	\\
			&	C2	&&	16:09:52.7	&	$-$51:54:53.9	&	1.6×1.1	&	7031	&	1530±160	&	6520±810	&	136±4	&	99.2±16.1	&	(1.2±0.2)e+24	&	1	&	H	\\
			&	C3	&&	16:09:52.4	&	$-$51:54:55.6	&	1.6×1.2	&	7344	&	1220±100	&	5500±550	&	110±2	&	105.0±15	&	(1.2±0.1)e+24	&	1	&	H	\\
			\hline	
			\multirow{1}{*}{I16071-5142} & C1 &	\multirow{1}{*}{5.3}
			&	16:10:59.8	&	$-$51:50:23.1	&	1.8×1.0	&	7111	&	849±107	&	3840±580	&	127±10	&	62.9±12.4	&	(7.5±1.3)e+23	&	1	&	H	\\
			\hline															
			\multirow{2}{*}{I16076-5134} & C1 &	\multirow{2}{*}{5.3}
			&	16:11:27.7	&	$-$51:41:55.6	&	0.9×0.7	&	4207	&	93±1	&	218±5	&	87±11	&	5.4±0.9	&	(1.8±0.2)e+23	&	1	&	I	\\
			&	C2	&&	16:11:26.5	&	$-$51:41:57.4	&	2.3×1.5	&	9844	&	160±11	&	1410±110	&	98±10	&	30.5±5	&	(1.9±0.2)e+23	&	1	&	H	\\
			\hline	
			\multirow{3}{*}{I16272-4837} & C1 &	\multirow{3}{*}{2.9}
			&	16:30:57.3	&	$-$48:43:40.1	&	1.0×0.8	&	2594	&	212±8	&	549±27	&	106±3	&	3.3±0.4	&	(2.9±0.2)e+23	&	0	&	I	\\
			&	C2	&&	16:30:58.6	&	$-$48:43:51.4	&	1.2×0.6	&	2461	&	252±14	&	638±49	&	102±17	&	4.0±0.8	&	(3.9±0.7)e+23	&	0	&	I	\\
			&	C3	&&	16:30:58.8	&	$-$48:43:54.0	&	0.8×0.8	&	2320	&	1144±51	&	2560±160	&	115±4	&	14.0±1.7	&	(1.6±0.1)e+24	&	1	&	H	\\
			\hline															
			\multirow{2}{*}{I16351-4722} & C1 &	\multirow{2}{*}{3.0}
			&	16:38:50.8	&	$-$47:27:54.1	&	0.8×0.6	&	2078	&	113±3	&	220±9	&	90±9	&	1.7±0.2	&	(2.3±0.3)e+23	&	0	&	L	\\
			&	C2	&&	16:38:50.5	&	$-$47:28:00.8	&	2.2×1.9	&	6134	&	320±45	&	3010±470	&	170±10	&	11.6±2.3	&	(1.9±0.3)e+23	&	1	&	H	\\
			\hline															
			\multirow{1}{*}{I17220-3609} & C1 &	\multirow{1}{*}{8.0}
			&	17:25:25.3	&	$-$36:12:44.1	&	3.6×1.4	&	17960	&	554±19	&	5870±220	&	107±13	&	263.1±42.6	&	(4.9±0.6)e+23	&	0	&	H	\\
			\hline
		\end{tabular}
		\begin{tablenotes}
			\item[ ] Notes. The distances of the 9 high-mass star-forming regions are taken from \cite{2020MNRAS.496.2790L}. The positions, deconvolved sizes, peak flux densities, and integrated flux densities of 19 dense cores are obtained from multi-component 2D Gaussian fitting of the 870 $\mu$m continuum with CASA. The radii are derived from the equation R$_{\rm core}$ = $\sqrt{\theta_{\rm maj} × \theta_{\rm min}}$ / 3600 × $\pi$ / 180 × D. The core masses and molecular hydrogen column densities are calculated in Eq. (\ref{eq1}) and Eq. (\ref{eq2}) of Section \ref{subsec:ce}. The mass classification is discussed in Section \ref{subsec:cc}.
			\item[ ] Outflows. 0 = the outflows are not detected, 1 = the outflows are detected (taken from \citealt{2020ApJ...890...44B}). 
		\end{tablenotes}
	\end{table}
\end{landscape}

\begin{landscape}
	\begin{table}
		\centering
		\caption{Physical Parameters of \MF, \DE, \K, $^{13}$CH$_{3}$OH, and CH$_{3}$OH.}
		\label{tab2}
		\begin{tabular}{lcccccccccccccc}
			\hline\hline
			\multirow{3}{*}{Region} & \multirow{3}{*}{Core} & \multicolumn{3}{c}{\MF} && \multicolumn{3}{c}{\DE} && \multicolumn{3}{c}{\K} & $^{13}$CH$_{3}$OH & CH$_{3}$OH \\
			\cline{3-5}\cline{7-9}\cline{11-13}
			&& T & N & $\Delta$V$_{\rm dec}$ && T & N & $\Delta$V$_{\rm dec}$ && T & N & $\Delta$V$_{\rm dec}$ & N & N \\
			&& (K) & (cm$^{-2}$) & (km s$^{-1}$) && (K) & (cm$^{-2}$) & (km s$^{-1}$) && (K) & (cm$^{-2}$) & (km s$^{-1}$) & (cm$^{-2}$) & (cm$^{-2}$) \\
			\hline
			\multirow{1}{*}{I14498-5856} 
			&	C1	&	102±6	&	(2.8±0.7)e+16	&	4.1±0.1	&&	65±4	&	(2.8±0.3)e+16	&	3.9±0.2	&&	105±2	&	(2.5±0.3)e+15	&	5.5±0.6	&	(2.5±0.9)e+16	&	(1.1±0.4)e+18$^{a}$	\\
			\hline
			\multirow{4}{*}{I15520-5234} 
			&	C1	&	80±10	&	(8.5±1.7)e+14	&	1.7±0.7	&&	66±14	&	(8.0±1.4)e+15	&	2.7±0.2	&&	85±39	&	(3.0±1.1)e+14	&	2.6±1.0	&	$\cdots$	&	(1.4±0.3)e+16	\\
			&	C2	&	70±36	&	(1.5±0.4)e+15	&	0.5±0.4	&&	63±10	&	(1.1±0.2)e+16	&	1.6±0.1	&&	110±23	&	(1.3±0.3)e+14	&	1.1±0.1	&	$\cdots$	&	(9.5±1.3)e+16	\\
			&	C3	&	102±4	&	(9.5±0.5)e+16	&	1.5±0.1	&&	80±1	&	(4.8±0.8)e+16	&	1.1±0.2	&&	105±8	&	(2.8±0.4)e+15	&	1.6±0.1	&	(9.5±0.7)e+16	&	(4.1±0.3)e+18$^{a}$	\\
			&	C4	&	105±37	&	(7.5±3.8)e+14	&	1.1±0.6	&&	77±38	&	(2.1±0.8)e+15	&	1.3±0.1	&&	88±46	&	(1.3±0.7)e+14	&	1.4±0.6	&	$\cdots$	&	(7.2±1.2)e+15	\\
			\hline
			\multirow{2}{*}{I15596-5301} 
			&	C1	&	100±23	&	(1.7±0.7)e+16	&	4.7±0.6	&&	58±8	&	(1.6±0.5)e+16	&	5.2±0.6	&&	105±14	&	(1.0±0.5)e+15	&	5.4±0.9	&	(5.7±1.1)e+15	&	(2.2±0.4)e+17$^{a}$	\\
			&	C2	&	110±39	&	(2.4±1.0)e+16	&	3.9±0.5	&&	60±6	&	(2.6±0.1)e+16	&	4.9±0.6	&&	127±19	&	(1.5±0.3)e+15	&	3.0±0.7	&	(1.6±0.4)e+16	&	(6.1±1.5)e+17$^{a}$	\\
			\hline
			\multirow{3}{*}{I16060-5146} 
			&	C1	&	112±7	&	(3.8±0.8)e+16	&	2.4±0.3	&&	65±2	&	(6.7±0.2)e+16	&	2.7±0.2	&&	136±12	&	(2.0±0.5)e+15	&	3.3±0.3	&	(3.2±0.3)e+16	&	(1.1±0.1)e+18$^{a}$	\\
			&	C2	&	136±4	&	(2.4±0.7)e+16	&	4.9±0.6	&&	137±5	&	(9.2±1.0)e+15	&	6.2±0.6	&&	173±1	&	(5.3±0.2)e+15	&	5.0±0.6	&	(3.7±0.9)e+16	&	(1.3±0.3)e+18$^{a}$	\\
			&	C3	&	110±2	&	(1.4±0.1)e+17	&	4.9±0.7	&&	80±13	&	(5.7±0.2)e+16	&	5.9±0.6	&&	147±4	&	(9.0±0.4)e+15	&	4.7±0.7	&	(1.2±0.1)e+17	&	(4.2±0.4)e+18$^{a}$	\\
			\hline																
			\multirow{1}{*}{I16071-5142} 
			&	C1	&	127±10	&	(2.0±0.1)e+17	&	7.4±0.5	&&	80±4	&	(1.8±0.1)e+17	&	6.2±0.7	&&	138±6	&	(1.6±0.1)e+16	&	7.9±0.8	&	(3.6±0.3)e+17	&	(1.3±0.1)e+19$^{a}$	\\
			\hline																
			\multirow{2}{*}{I16076-5134} 
			&	C1	&	87±11	&	(3.6±0.3)e+15	&	2.6±0.8	&&	58±10	&	(3.1±0.6)e+15	&	2.8±0.9	&&	130±8	&	(2.0±0.2)e+14	&	3.4±0.9	&	(2.0±0.3)e+15	&	(7.0±0.1)e+16$^{a}$	\\
			&	C2	&	98±10	&	(3.8±0.3)e+16	&	4.5±0.5	&&	51±3	&	(5.8±0.4)e+16	&	6.4±0.6	&&	135±9	&	(2.3±0.5)e+15	&	5.4±0.7	&	(3.6±0.6)e+16	&	(1.3±0.2)e+18$^{a}$	\\
			\hline															
			\multirow{3}{*}{I16272-4837} 
			&	C1	&	106±3	&	(1.0±0.1)e+17	&	4.4±0.5	&&	79±2	&	(9.0±0.3)e+16	&	4.4±0.5	&&	103±3	&	(1.0±0.1)e+16	&	4.9±0.5	&	(9.0±2.9)e+16	&	(3.7±1.2)e+18$^{a}$	\\
			&	C2	&	102±17	&	(5.0±0.7)e+16	&	5.6±0.6	&&	62±1	&	(3.4±0.8)e+16	&	4.9±0.5	&&	135±10	&	(3.5±0.7)e+15	&	7.4±0.7	&	(7.2±0.7)e+16	&	(3.0±0.3)e+18$^{a}$	\\
			&	C3	&	115±4	&	(7.0±0.9)e+17	&	2.8±0.5	&&	109±5	&	(3.5±0.5)e+17	&	4.5±0.6	&&	125±7	&	(4.8±0.7)e+16	&	3.7±0.7	&	(5.5±0.2)e+17	&	(2.3±0.1)e+19$^{a}$	\\
			\hline																
			\multirow{2}{*}{I16351-4722} 
			&	C1	&	90±9	&	(1.0±0.1)e+16	&	1.6±0.6	&&	50±4	&	(7.2±0.1)e+15	&	1.6±0.7	&&	123±19	&	(4.8±0.8)e+14	&	2.1±0.8	&	(7.0±0.5)e+15	&	(2.9±0.2)e+17$^{a}$	\\
			&	C2	&	170±10	&	(2.0±0.1)e+17	&	5.3±0.6	&&	93±5	&	(1.3±0.2)e+17	&	4.9±0.6	&&	159±33	&	(1.3±0.5)e+16	&	5.9±0.8	&	(1.6±0.2)e+17	&	(6.6±0.8)e+18$^{a}$	\\
			\hline															
			\multirow{1}{*}{I17220-3609} 
			&	C1	&	107±13	&	(1.2±0.2)e+17	&	3.9±0.5	&&	65±15	&	(1.5±0.1)e+17	&	5.9±0.5	&&	102±10	&	(1.2±0.1)e+16	&	3.8±0.6	&	(6.5±1.1)e+16	&	(1.2±0.2)e+18$^{a}$	\\
			\hline
		\end{tabular}
		\begin{tablenotes}
			\item[ ] Notes. The rotation temperatures T and column densities N are model fitting results. The $\Delta$V$_{\rm dec}$ is deconvolved line widths. $^{a}$ The column densities of CH$_{3}$OH in these cores are obtained from the column densities of $^{13}$CH$_{3}$OH using Eq. (\ref{eq7}).
		\end{tablenotes}
	\end{table}
\end{landscape}

\begin{itemize}
	\item[1.] Most transitions in each core are optically thin ($\tau$ $<$ 1).
	
	\item[2.] In each core, more than three lines are detected for \MF, \DE, and \K, and \MF\ presents more lines than \DE\ and \K.
	
	\item[3.] In all cores, the \MF\ and \DE\ have larger line intensities, while the line intensities of \K\ are generally smaller.
\end{itemize}
The rotational transitions of \MF, \DE, and \K\ detected in 19 dense cores are listed in Table \ref{tab4} of Appendix \ref{sec:mt}. The upper level energy ranges of \MF, \DE, and \K\ are 80 to 590 K, 73 to 167 K and 148 to 474 K, respectively. These three molecules, especially \MF\ and \K, cover a wide range of upper level energies, which is conducive to the constraints of rotational temperatures and column densities.

\subsection{Column Densities, Rotation Temperatures, Line Widths, and Molecular Abundances \label{subsec:cdrtlwama}}

The MAGIX optimization results for \MF, \DE, and \K\ in 19 dense cores are shown in Table \ref{tab2}. The column densities mainly range from 10$^{15}$ to 10$^{17}$ cm$^{-2}$ for \MF\ and \DE, and from 10$^{14}$ to 10$^{16}$ cm$^{-2}$ for \K. The upper level energies of \K\ lines are larger than 148 K (see Section \ref{subsec:li}), and the column densities of \K\ are lower, which can explain its weaker line intensities than the other two molecules. The rotation temperature ranges of \MF, \DE, and \K\ are 70 to 170 K, 50 to 137 K and 85 to 173 K, respectively. The rotation temperatures of \DE\ in most cores are generally lower than those of \MF\ and \K, possibly because our 870 $\mu$m observations of the \DE\ lines have lower upper level energies (see Table \ref{tab4}) and then the hot components of the cores are not sampled. \MF\ and \K\ should trace hotter components than \DE. Considering the overestimation of line widths due to velocity resolution of 0.98 km s$^{-1}$ (see Section \ref{sec:Observations}), we adopt the deconvolved line widths by the following formula:
\begin{equation} \label{eq4}
	\rm \Delta V_{dec} = \sqrt{\Delta V^{2} - \Delta v^{2}},
\end{equation}
where $\Delta$V is the fitted line width convolved with the velocity resolution $\Delta$v. We computed the molecular abundances relative to H$_{2}$ and CH$_{3}$OH by the following formula:
\begin{equation} \label{eq5}
	\rm f_{H_2} = N / N (H_{2}),
\end{equation}
\begin{equation} \label{eq6}
	\rm f_{CH_3OH} = N / N (CH_{3}OH),
\end{equation}
where N is the column density of the specific molecule, N (H$_{2}$) is the column density of H$_{2}$ (see Table \ref{tab1}), and N (CH$_{3}$OH) is the column density of CH$_{3}$OH (see Table \ref{tab2}). Due to the blending of CH$_{3}$OH lines with other molecular lines in 16 line-rich cores, we then fit $^{13}$CH$_{3}$OH line transitions and derive the column densities of CH$_{3}$OH by the column densities of $^{13}$CH$_{3}$OH (see Table \ref{tab2}) multiplied by the $^{12}$C/$^{13}$C ratios from the following formula \citep{2019ApJ...877..154Y}:
\begin{equation} \label{eq7}
	\rm ^{12}C / ^{13}C = (5.08 \pm 1.10)R_{GC} + (11.86 \pm 6.60),
\end{equation}
where R$_{\rm GC}$ (in kpc) represents the distance from the Galactic Center \citep{2020MNRAS.496.2790L}. In the other 3 line-rich cores without $^{13}$CH$_{3}$OH detected, the column densities of CH$_{3}$OH are obtained through direct fitting, where the CH$_{3}$OH lines are not blended. The abundances of \MF, \DE, and \K\ relative to H$_{2}$ and CH$_{3}$OH are listed in Table \ref{tab3}.

\begin{table*}
	\centering
	\caption{Molecular Abundances Relative to H$_{2}$ and CH$_{3}$OH.} 
	\label{tab3}
	\begin{tabular}{lccccccccc}
		\hline\hline
		\multirow{2}{*}{Region} & \multirow{2}{*}{Core} & \multicolumn{2}{c}{\MF} && \multicolumn{2}{c}{\DE} && \multicolumn{2}{c}{\K} \\
		\cline{3-4}\cline{6-7}\cline{9-10}
		&& f$_{\rm H_2}$ & f$_{\rm CH_3OH}$ && f$_{\rm H_2}$ & f$_{\rm CH_3OH}$ && f$_{\rm H_2}$ & f$_{\rm CH_3OH}$ \\
		\hline
		\multirow{1}{*}{I14498-5856} 
		&	C1	&	(1.4±0.4)e$-$07	&	(2.5±1.1)e$-$02	&&	(1.4±0.2)e$-$07	&	(2.5±1.0)e$-$02	&&	(1.2±0.2)e$-$08	&	(2.3±0.9)e$-$03	\\
		\hline
		\multirow{4}{*}{I15520-5234} 
		&	C1	&	(5.3±1.3)e$-$09	&	(6.1±1.8)e$-$02	&&	(5.0±1.1)e$-$08	&	(5.7±1.6)e$-$01	&&	(1.9±0.7)e$-$09	&	(2.1±0.9)e$-$02	\\
		&	C2	&	(1.3±0.8)e$-$08	&	(1.6±0.5)e$-$02	&&	(9.9±5.4)e$-$08	&	(1.2±0.3)e$-$01	&&	(1.2±0.7)e$-$09	&	(1.4±0.4)e$-$03	\\
		&	C3	&	(5.3±0.4)e$-$07	&	(2.3±0.2)e$-$02	&&	(2.7±0.5)e$-$07	&	(1.2±0.2)e$-$02	&&	(1.6±0.2)e$-$08	&	(6.9±1.1)e$-$04	\\
		&	C4	&	(6.8±4.3)e$-$09	&	(1.0±0.6)e$-$01	&&	(1.9±1.0)e$-$08	&	(2.9±1.2)e$-$01	&&	(1.2±0.8)e$-$09	&	(1.8±1.0)e$-$02	\\
		\hline
		\multirow{2}{*}{I15596-5301} 
		&	C1	&	(2.4±1.2)e$-$07	&	(7.8±3.5)e$-$02	&&	(2.3±0.9)e$-$07	&	(7.4±2.7)e$-$02	&&	(1.4±0.8)e$-$08	&	(4.6±2.5)e$-$03	\\
		&	C2	&	(1.4±0.8)e$-$07	&	(3.9±1.9)e$-$02	&&	(1.6±0.6)e$-$07	&	(4.3±1.1)e$-$02	&&	(9.1±3.7)e$-$09	&	(2.5±0.8)e$-$03	\\
		\hline
		\multirow{3}{*}{I16060-5146} 
		&	C1	&	(5.0±1.1)e$-$08	&	(3.4±0.8)e$-$02	&&	(8.8±0.8)e$-$08	&	(6.0±0.6)e$-$02	&&	(2.6±0.7)e$-$09	&	(1.8±0.5)e$-$03	\\
		&	C2	&	(2.0±0.6)e$-$08	&	(1.9±0.7)e$-$02	&&	(7.6±1.3)e$-$09	&	(7.1±1.8)e$-$03	&&	(4.4±0.6)e$-$09	&	(4.1±1.0)e$-$03	\\
		&	C3	&	(1.2±0.2)e$-$07	&	(3.3±0.4)e$-$02	&&	(4.8±0.5)e$-$08	&	(1.4±0.1)e$-$02	&&	(7.7±0.9)e$-$09	&	(2.1±0.2)e$-$03	\\
		\hline																
		\multirow{1}{*}{I16071-5142} 
		&	C1	&	(2.7±0.5)e$-$07	&	(1.5±0.1)e$-$02	&&	(2.4±0.4)e$-$07	&	(1.4±0.1)e$-$02	&&	(2.1±0.4)e$-$08	&	(1.2±0.1)e$-$03	\\
		\hline																
		\multirow{2}{*}{I16076-5134} 
		&	C1	&	(2.0±0.3)e$-$08	&	(5.1±1.0)e$-$02	&&	(1.7±0.4)e$-$08	&	(4.4±1.1)e$-$02	&&	(1.1±0.2)e$-$09	&	(2.9±0.6)e$-$03	\\
		&	C2	&	(2.0±0.3)e$-$07	&	(2.9±0.5)e$-$02	&&	(3.1±0.4)e$-$07	&	(4.5±0.8)e$-$02	&&	(1.2±0.3)e$-$08	&	(1.8±0.5)e$-$03	\\
		\hline															
		\multirow{3}{*}{I16272-4837} 
		&	C1	&	(3.4±0.4)e$-$07	&	(2.7±0.9)e$-$02	&&	(3.1±0.2)e$-$07	&	(2.4±0.8)e$-$02	&&	(3.4±0.4)e$-$08	&	(2.7±0.9)e$-$03	\\
		&	C2	&	(1.3±0.3)e$-$07	&	(1.7±0.3)e$-$02	&&	(8.6±2.6)e$-$08	&	(1.1±0.3)e$-$02	&&	(8.9±2.4)e$-$09	&	(1.2±0.3)e$-$03	\\
		&	C3	&	(4.5±0.7)e$-$07	&	(3.1±0.4)e$-$02	&&	(2.2±0.4)e$-$07	&	(1.6±0.2)e$-$02	&&	(3.1±0.5)e$-$08	&	(2.1±0.3)e$-$03	\\
		\hline																
		\multirow{2}{*}{I16351-4722} 
		&	C1	&	(4.3±0.6)e$-$08	&	(3.5±0.4)e$-$02	&&	(3.1±0.3)e$-$08	&	(2.5±0.2)e$-$02	&&	(2.1±0.4)e$-$09	&	(1.7±0.3)e$-$03	\\
		&	C2	&	(1.1±0.2)e$-$06	&	(3.0±0.4)e$-$02	&&	(7.0±1.6)e$-$07	&	(2.0±0.4)e$-$02	&&	(7.0±2.9)e$-$08	&	(2.0±0.8)e$-$03	\\
		\hline															
		\multirow{1}{*}{I17220-3609} 
		&	C1	&	(2.4±0.5)e$-$07	&	(1.0±0.2)e$-$01	&&	(3.0±0.4)e$-$07	&	(1.3±0.2)e$-$01	&&	(2.4±0.4)e$-$08	&	(1.0±0.2)e$-$02	\\
		\hline
	\end{tabular}
	\begin{tablenotes}
		\item[ ] Notes. The abundances of \MF, \DE, and \K\ relative to H$_{2}$ and CH$_{3}$OH are calculated in Eq. (\ref{eq5}) and Eq. (\ref{eq6}) of Section \ref{subsec:cdrtlwama}.
	\end{tablenotes}
\end{table*}

\subsection{The Molecular Emission Maps \label{subsec:tmem}}
The integrated intensity maps of \MF\ at 342572 MHz (E$_{\rm u}$ = 81 K), \DE\ at 344358 MHz (E$_{\rm u}$ = 167 K), and \K\ at 343173 MHz (E$_{\rm u}$ = 148 K) in 9 high-mass star-forming regions are shown in Figure \ref{fig2}. The emission peaks of \MF, \DE, and \K\ are consistent, and the spatial distribution of the three molecules is similar. These results suggest that there may be physical or chemical links among the three molecules. In addition, there is no difference in the spatial distributions of different energy level transitions of \MF\ in the 9 high-mass star-forming regions, as shown in Figure \ref{fig9} of Appendix \ref{sec:sd}. From Figure \ref{fig2} and Figure \ref{fig9}, the line emissions of \MF, \DE, and \K\ are primarily distributed around intense continuum emission. I15520, I16060, and I17220 are found to be associated with intense UC H{\sc ii} regions, while I16071, I16076, and I16351 are associated with weaker UC H{\sc ii} regions \citep{2022MNRAS.511.3463Q, 2023MNRAS.520.3245Z}. In I15520, I16060, and I17220, the line emissions are offset from the continuum cores, and \MF, \DE, and \K\ also show irregular emissions in these regions, likely due to the influence of UC H{\sc ii} regions.

\section{Discussion} \label{sec:Discussion}

\subsection{Core Classification \label{subsec:cc}}
Unlike HMCs, which are associated with the formation of massive stars, hot corinos are found around low-mass protostars. Typical hot corinos are small in size ($\lesssim$ 200AU) and show a rich chemistry \citep{2003ApJ...593L..51C, 2004ApJ...615..354B, 2004A&A...416..577M, 2007A&A...463..601B}. Intermediate-mass hot cores (IMHCs) provide the link between HMCs and hot corinos, although it has rarely been reported \citep{2010ApJ...721L.107S, 2011ApJ...743L..32P, 2014A&A...568A..65F}. In high-mass star-forming regions, dense cores exist within massive reservoirs, and the mass of the stars formed is uncertain. Therefore, it is difficult to identify hot corinos and IMHCs (sometimes they are simply assumed to be HMCs) in high-mass star-forming regions. 

In Table \ref{tab1}, we classify 19 line-rich cores into three groups according to their mass (further material accretion or loss may occur), namely 12 high-mass line-rich cores (H; $>$ 8 M$_{\bigodot}$), 6 intermediate-mass line-rich cores (I; 2 – 8 M$_{\bigodot}$), and 1 low-mass line-rich core (L; $<$ 2 M$_{\bigodot}$). Moreover, the 11 high-mass star-forming regions are subsamples of \cite{2022MNRAS.511.3463Q}, who reported 60 HMCs from 146 high-mass star-forming regions at ALMA resolutions of $\sim$ 1.2 – 1.9 arcsec. Our currently higher spatial resolution observations reveal that part of the HMCs detected in previous works actually host multiple line-rich cores with different masses. (see \citealt{2022MNRAS.511.3463Q} and our Figure \ref{fig2}).

\subsection{Abundance, Temperature and Line Width Correlations \label{subsec:atalwc}}
\begin{figure}
	\includegraphics[width=\linewidth]{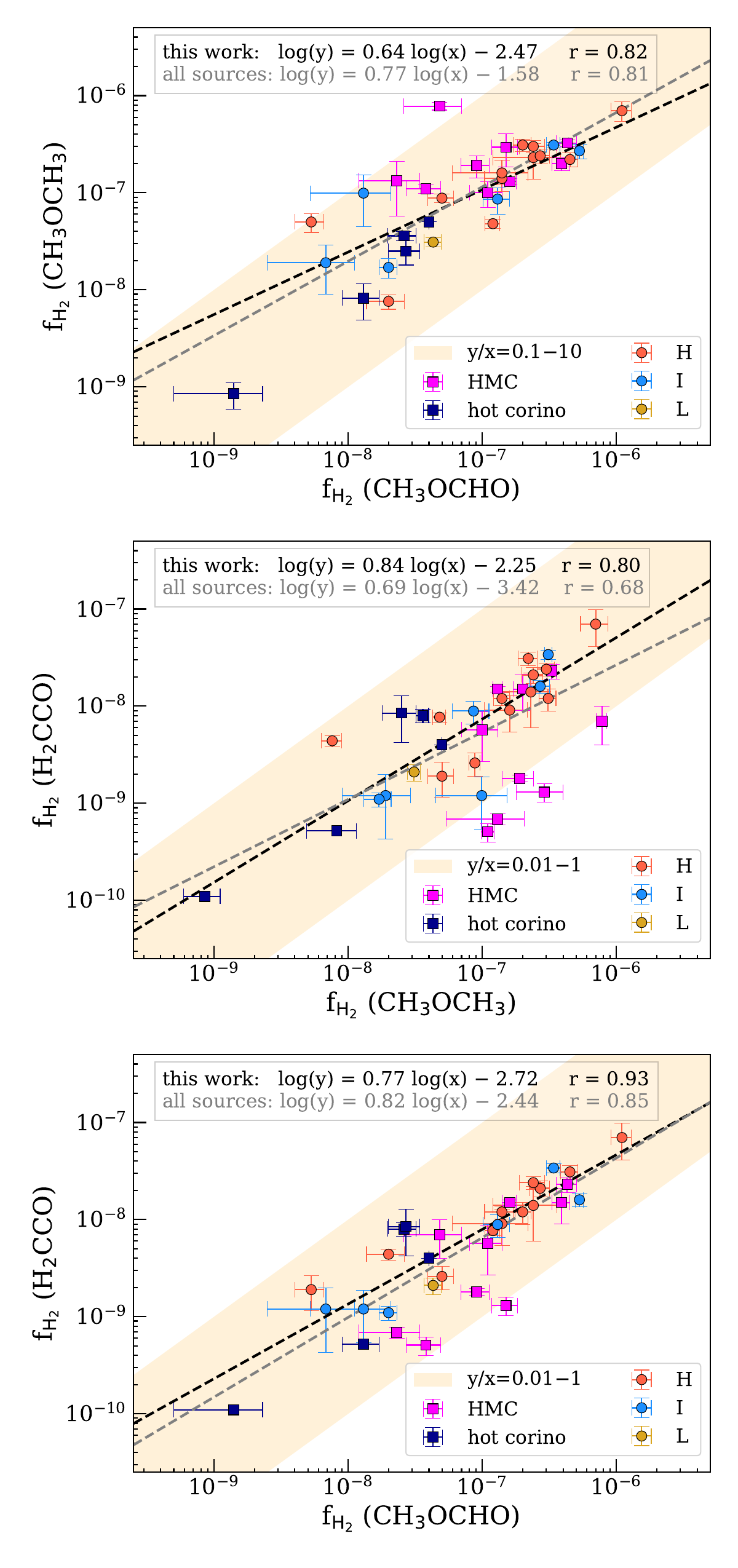}
	\caption{Comparison among the molecular abundances of \MF, \DE, and \K\ relative to H$_{2}$. The colored circles represent the different types of cores in this work (H = high-mass line-rich core, I = intermediate-mass line-rich core, and L = low-mass line-rich core), while the colored squares represent the sources in the literature (References: HMCs: \citealt{2015A&A...581A..71F, 2019A&A...628A...2B, 2022MNRAS.512.4419P}; hot corinos: \citealt{2015ApJ...804...81T, 2017MNRAS.469L..73L, 2018A&A...618A.145O, 2019MNRAS.483.1850B}). The orange areas indicate the relative abundance ranges of molecules. The black dashed lines and the gray dashed lines are linear least-squares fits of this work and all sources, respectively. The fitting results and Pearson correlation coefficients (r) are shown on the top left.} \label{fig3}
\end{figure}

\begin{figure}
	\includegraphics[width=\linewidth]{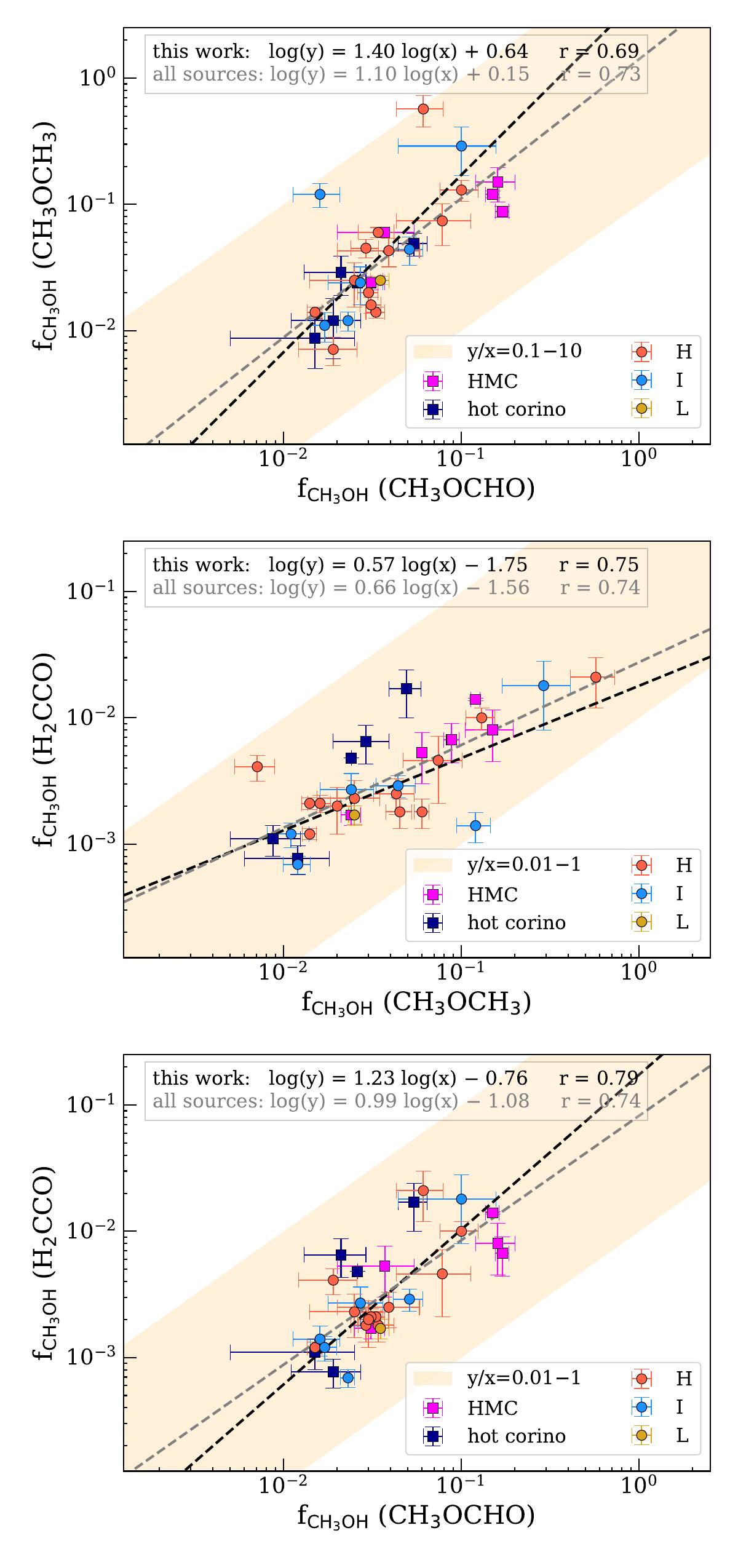}
	\caption{Comparison among the molecular abundances of \MF, \DE, and \K\ relative to CH$_{3}$OH. The colored circles represent the different types of cores in this work (H = high-mass line-rich core, I = intermediate-mass line-rich core, and L = low-mass line-rich core), while the colored squares represent the sources in the literature (References: HMCs: \citealt{2019A&A...628A...2B, 2022MNRAS.512.4419P}; hot corinos: \citealt{2015ApJ...804...81T, 2017MNRAS.469L..73L, 2018A&A...618A.145O, 2018A&A...620A.170J}). The orange areas indicate the relative abundance ranges of molecules. The black dashed lines and the gray dashed lines are linear least-squares fits of this work and all sources, respectively. The fitting results and Pearson correlation coefficients (r) are shown on the top left.} \label{fig4}
\end{figure}

In Figure \ref{fig3}, we compare the \MF, \DE, and \K\ abundances relative to H$_{2}$ in 19 dense cores. \MF\ and \DE\ show a strong abundance correlation (the Pearson correlation coefficient r = 0.82). The abundance correlation between \DE\ and \K\ is significant (r = 0.80). \MF\ and \K\ show the stronger abundance correlation (r = 0.93) than the other two pairs of molecules. Figure \ref{fig4} shows the relationships of the abundances of the three molecules relative to CH$_{3}$OH in 19 dense cores. The abundances of \MF\ and \DE\ relative to CH$_{3}$OH mainly range from 10$^{-2}$ to 10$^{-1}$, while the abundance of \K\ relative to CH$_{3}$OH mainly ranges from 10$^{-3}$ to 10$^{-2}$. The abundances of the three molecules relative to CH$_{3}$OH in our observations are consistent with those observed in high-mass star-forming regions \citep{2019A&A...628A...2B, 2022MNRAS.512.4419P, 2023A&A...678A.137C, 2024MNRAS.529.3244L}, intermediate-mass star-forming regions \citep{2011ApJ...743L..32P, 2014A&A...568A..65F, 2018A&A...618A.145O}, low-mass star-forming regions \citep{2015ApJ...804...81T, 2017MNRAS.469L..73L, 2018A&A...620A.170J}, Galactic center molecular clouds \citep{2006A&A...455..971R, 2008ApJ...672..352R}, and comets \citep{2019ESC.....3.1550B}. The correlation coefficients of the abundances relative to CH$_{3}$OH are 0.69 between \MF\ and \DE, 0.75 between \DE\ and \K, and 0.79 between \MF\ and \K, indicating positive correlations. The obvious abundance correlation relative to CH$_{3}$OH between \MF\ and \DE\ was also observed in four different interstellar sources, including 1 HMC, 1 hot corino, and 2 comets \citep{2024MNRAS.529.3244L}. From Figure \ref{fig3} and Figure \ref{fig4}, the relative abundances of \MF\ to \DE\ are almost constant equal to 1. \cite{2023A&A...678A.137C} also found a constant relative abundance of $\sim$ 1 between \MF\ and \DE\  toward 19 protostars with different luminosities. This ratio is consistent with the observations from low-, intermediate- and high-mass star-forming regions \citep{2014ApJ...791...29J, 2017A&A...598A..59R, 2018A&A...618A.145O, 2020A&A...641A..54C, 2022MNRAS.512.4419P}. In Figure \ref{fig3} and Figure \ref{fig4}, we also compared our results with the molecular abundances relative to H$_{2}$ and CH$_{3}$OH in other HMCs and hot corinos. The abundance correlations relative to H$_{2}$ and CH$_{3}$OH in these sources agree with the results in our observations but more dispersed than our samples, which is most likely due to different spatial resolution and spectral setup of these observations. These results suggest that the formation of the three molecules may be closely related, i.e., they may have similar production conditions, or even share a common chemical reaction network.

\begin{figure}
	\includegraphics[width=\linewidth]{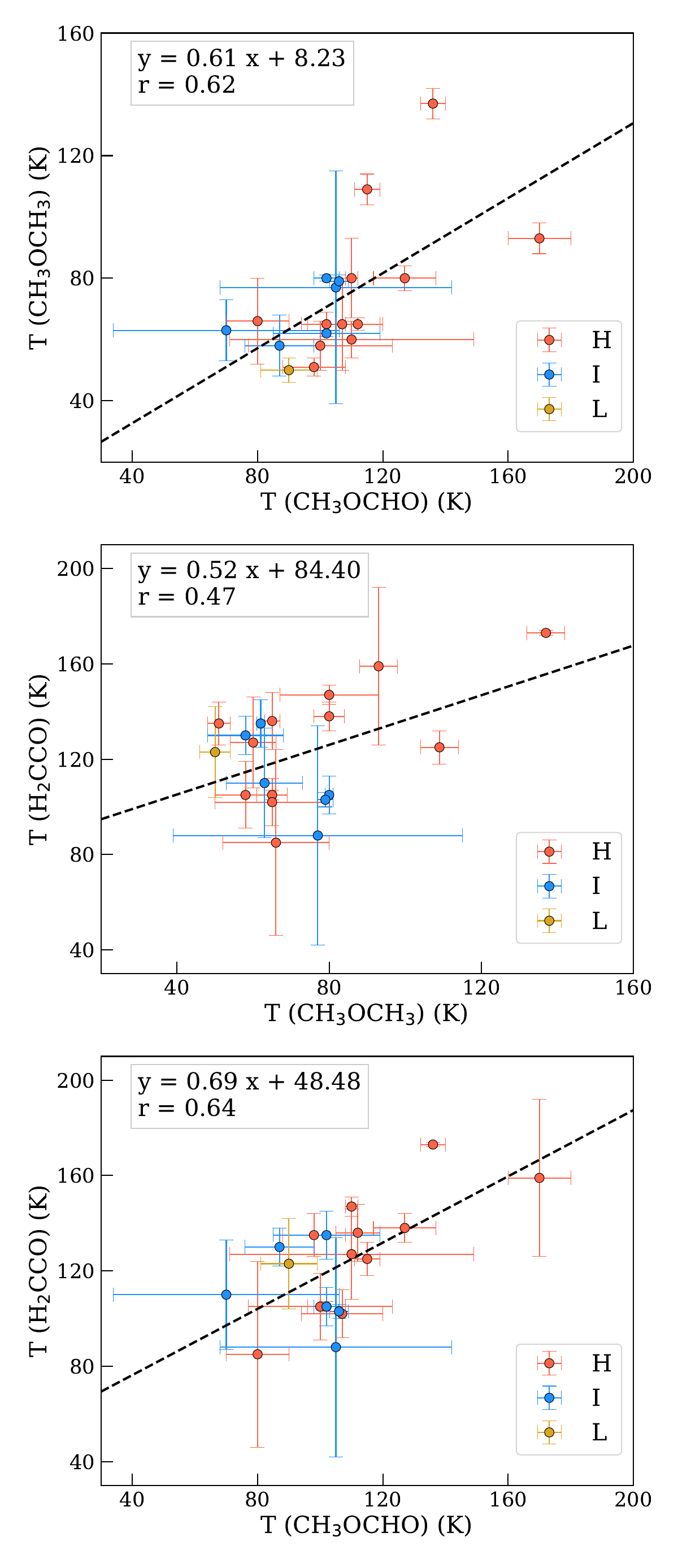}
	\caption{Comparison among the rotation temperatures of \MF, \DE, and \K. The colored circles represent different types of cores (H = high-mass line-rich core, I = intermediate-mass line-rich core, and L = low-mass line-rich core). The black dashed lines are the linear least-squares fits to the data. The fitting results and Pearson correlation coefficients (r) are shown on the top left.} \label{fig5}
\end{figure}

In Figure \ref{fig5}, we compare the rotation temperatures of the three molecules. The temperature correlation coefficient of \MF\ and \DE\ is 0.62, \DE\ and \K\ is 0.47, and \MF\ and \K\ is 0.64. The results indicate that the rotation temperatures have no significant correlations among the three molecules. The results observed by \cite{2020A&A...641A..54C} toward 13 high-mass star-forming regions also showed a poor temperature correlation (r = 0.45) between \MF\ and \DE, but the overall temperature ranges of the two molecules are similar. In fact, most of the temperatures of the three molecules in our case are in the range of tens of Kelvin, which is probably the reason why there is no apparent trend in temperature.

\begin{figure}
	\includegraphics[width=\linewidth]{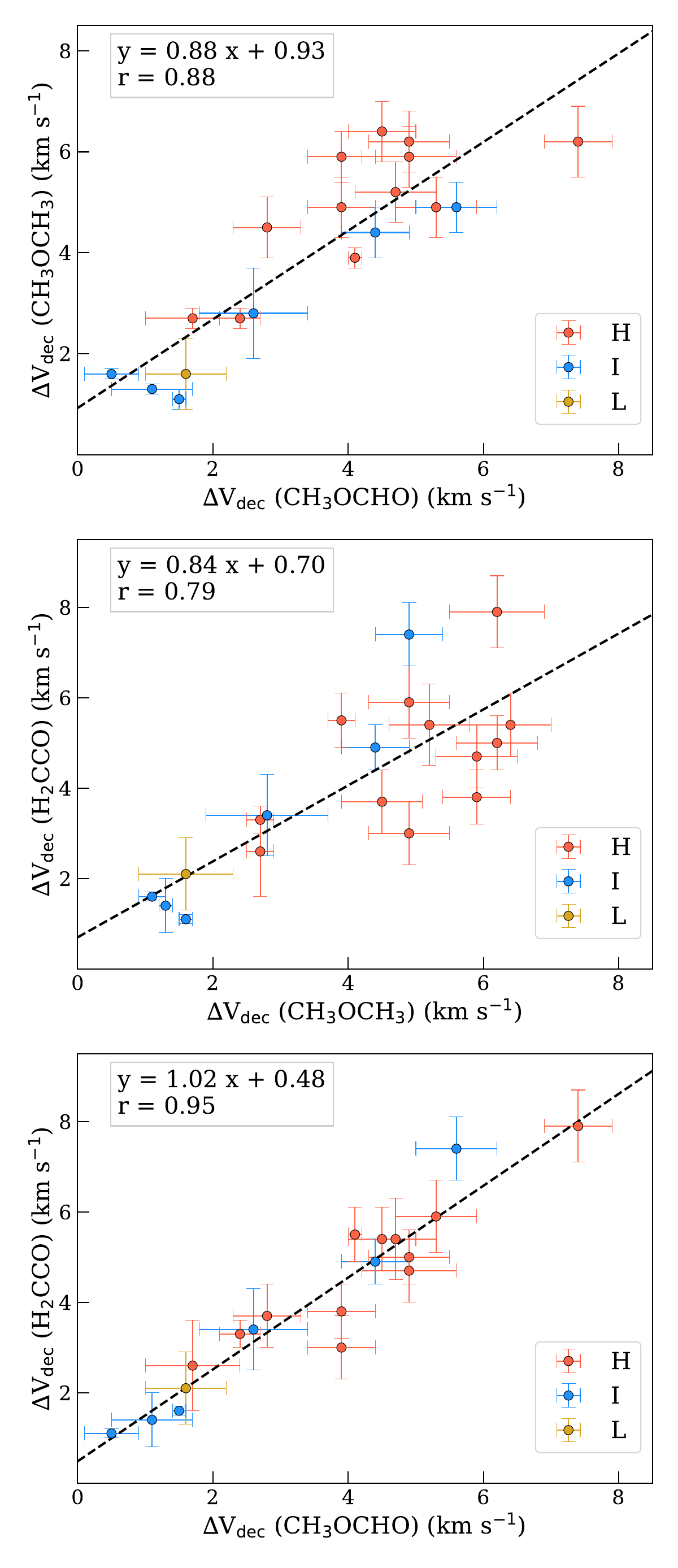}
	\caption{Comparison among the line widths of \MF, \DE, and \K. The colored circles represent different types of cores (H = high-mass line-rich core, I = intermediate-mass line-rich core, and L = low-mass line-rich core). The black dashed lines are the linear least-squares fits to the data. The fitting results and Pearson correlation coefficients (r) are shown on the top left.} \label{fig6}
\end{figure}

\begin{figure*}
	\includegraphics[width=\linewidth]{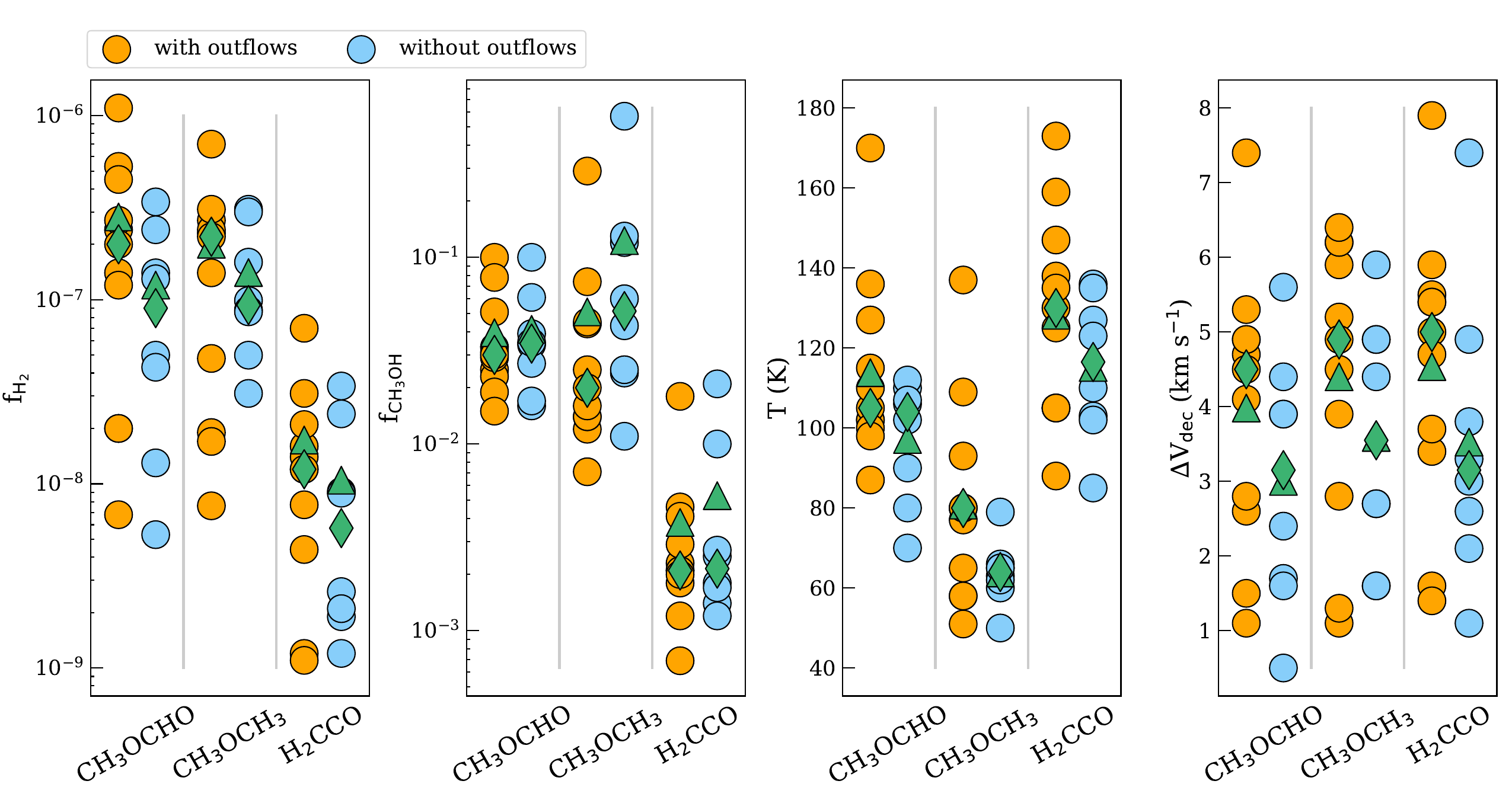}
	\caption{Comparison of molecular abundances relative to H$_{2}$ and CH$_{3}$OH, rotation temperatures and line widths of the cores with and without detected outflows. The colored circles represent cores with and without outflows. The green triangles represent the averages and the green diamonds represent the medians.} \label{fig7}
\end{figure*}

The line width relationships of the three molecules are shown in Figure \ref{fig6}. The \MF\ and \K\ show strong line width correlation (r = 0.95), followed by \MF\ and \DE\ (r = 0.88), and by \DE\ and \K\ (r = 0.79). This agrees with the line width correlation of \MF\ and \DE\ observed in 13 high-mass star-forming regions \citep{2020A&A...641A..54C}. This implies that the three molecules may trace the similar kinematics.\footnote{In our case, the line widths of the three molecules are almost entirely contributed by non-thermal motions: $\Delta$ V$_{\rm dec}$ $\sim$ $\Delta$ V$_{\rm NT}$ = $\sqrt{\Delta \rm V_{\rm dec}^{2} - 8ln2\frac{\rm kT_{\rm ex}}{\rm m}}$, where $\Delta$V$_{\rm NT}$ is the line width contributed by non-thermal motions, k is the Boltzmann constant, T$_{\rm ex}$ is the excitation temperature of the molecule, and m is the molecular mass.}

In Figures \ref{fig3}, \ref{fig5}, and \ref{fig6}, the H cores are shifted to the upper right relative to the I and L cores, indicating a tendency for the abundances relative to H$_{2}$, temperatures and line widths of the three molecules to be higher in massive line-rich cores. In Figure \ref{fig7}, we also compare molecular abundances relative to H$_{2}$ and CH$_{3}$OH, rotation temperatures and line widths of the cores with and without detected outflows (see Table \ref{tab1}). The abundances relative to H$_{2}$, temperatures and line widths of the three molecules tend to be higher where outflows are detected. These results confirm that massive cores and cores with outflows may have richer chemistry \citep{2020ARA&A..58..727J, 2023ASPC..534..379C}.

\subsection{Implications for Chemistry \label{subsec:iftc}}

The spatial similarities, abundance correlations and line width correlations of \MF, \DE, and \K\ are obvious. We briefly summarize the chemical models of these three molecules below. In gas-phase chemistry, protonated methanol (CH$_{3}$OH$_{2}^{+}$) can react with CH$_{3}$OH to produce \DE\  as follows \citep{1995ApJ...448..232C, 2016ApJ...821...46T, 2020ARA&A..58..727J}: 
\begin{equation} \label{eq8}
	\rm CH_3OH_2^+ \stackrel{CH_3OH}{\longrightarrow} CH_3OCH_4^+ \stackrel{e^- / NH_3}{\longrightarrow} CH_3OCH_3.
\end{equation}
\cite{2015MNRAS.449L..16B} proposed that \DE\ could generate \MF\ in cold gas environments through the following reactions:
\begin{equation} \label{eq9}
	\rm CH_3OCH_3 \stackrel{F / Cl}{\longrightarrow} CH_3OCH_2 \stackrel{O}{\longrightarrow} CH_3OCHO,
\end{equation}
where \DE\ is a precursor to \MF. In the case of grain-surface chemistry, the \MF\ and \DE\ follow the reactions \citep{2006A&A...457..927G, 2008ApJ...682..283G}:
\begin{equation} \label{eq10}
	\rm CH_3O + HCO \longrightarrow CH_3OCHO,
\end{equation}
\begin{equation} \label{eq11}
	\rm CH_3O + CH_3 \longrightarrow CH_3OCH_3,
\end{equation}
where methoxide (CH$_{3}$O) is the common precursor of \MF\ and \DE. On the surface of the dust grains, the aldehyde (HCO) can produce H$_{2}$CCO as follows \citep{2001AcSpA..57..685C, 2005IAUS..231..237C, 2017ApJ...847...89K}:
\begin{equation} \label{eq12}
	\rm HCO \stackrel{C}{\longrightarrow} HCCO \stackrel{H}{\longrightarrow} H_2CCO,
\end{equation}
\begin{equation} \label{eq13}
	\rm CH_2 + CO \longrightarrow H_2CCO,
\end{equation}
where HCO is the common precursor of \MF\ and \K\ through Eq. (\ref{eq10}) (\ref{eq12}). The spatial similarities, abundance correlations, and line width correlations of \MF, \DE, and \K, combined with chemical models, suggest that three molecules are chemically related. \cite{2016ApJ...821...46T} predicted the abundances of \DE\ relative to H$_{2}$ to be $\sim$ 10$^{-7}$ and relative to CH$_{3}$OH to be $\sim$ 10$^{-2}$ using reaction (\ref{eq8}). Our observed abundances of \DE\ relative to H$_{2}$ and CH$_{3}$OH are consistent with this prediction. \cite{2022ApJS..259....1G} showed a chemical network that includes reactions  (\ref{eq8}), (\ref{eq9}), (\ref{eq10}), (\ref{eq11}), and (\ref{eq13}) to predict the abundances of \MF, \DE, and \K. The three warm-up timescales of 5 × 10$^{4}$ years (fast), 2 × 10$^{5}$ years (medium), and 1 × 10$^{6}$ years (slow) were employed in their models. At the medium timescale, the predicted abundances relative to H$_{2}$ are $\sim$ 10$^{-7}$ for \MF\ and \DE, $\sim$ 10$^{-8}$ for \K, and relative to CH$_{3}$OH are $\sim$ 10$^{-2}$ for \MF\ and \DE, $\sim$ 10$^{-3}$ for \K. Our observed abundances of the three molecules relative to H$_{2}$ and CH$_{3}$OH are consistent with the medium timescale predictions. In conclusion, our observed abundances of the three molecules support both grain-surface and gas-phase chemical pathways for the production of \MF\ and \DE, and grain-surface chemical pathways for the production of \K.

\section{Conclusions} \label{sec:Conclusions}
We have analyzed the spectra of 11 high-mass star-forming regions obtained by ALMA band 7 observations, and studied the correlations of complex organic molecules \MF\ and \DE\ as well as an important precursor of complex organic molecules \K. We summarize the main results in the following:
\begin{itemize}
	\item[1.] \MF, \DE, and \K\ lines were detected in 19 line-rich cores from 9 out of 11 high-mass star-forming regions. At our higher spatial resolution observations, some of the hot molecular cores found in previous observations are revealed to actually host multiple line-rich cores with different masses.
	
	\item[2.] The integrated intensity maps of the 9 high-mass star-forming regions show that the emission peaks of the three molecules are consistent, and the spatial distribution of the molecules is similar. The emissions of the three molecules in the 9 high-mass star-forming regions are primarily distributed around intense continuum emission.
	
	\item[3.] The abundances relative to H$_{2}$ and CH$_{3}$OH, and line widths of the three molecules show obvious correlations in the 19 dense cores. The abundance correlations of the three molecules relative to H$_{2}$ and CH$_{3}$OH in other hot molecular cores and hot corinos agree with the results in our observations. The molecular abundances of \MF\ and \DE\ are rather similar, while the molecular abundances of \K\ are one order of magnitude lower.
	
	\item[4.] The abundances relative to H$_{2}$, temperatures and line widths of the three molecules tend to be higher in the cores with higher mass and with outflows. This confirms that massive cores and cores with outflows may have richer chemistry.
	
	\item[5.] The spatial similarities, abundance correlations, and line width correlations of \MF, \DE, and \K, combined with chemical models, suggest that three molecules are chemically related. Our results suggest that both grain-surface and gas-phase chemical pathways can be responsible for producing \MF\ and \DE, while \K\ should be produced by grain-surface chemical pathways.
\end{itemize}

\section*{Acknowledgements}
	This paper makes use of the following ALMA data: ADS/JAO.ALMA\#2017.1.00545.S. ALMA is a partnership of ESO (representing its member states), NSF (USA), and NINS (Japan), together with NRC (Canada), MOST and ASIAA (Taiwan), and KASI (Republic of Korea), in cooperation with the Republic of Chile. The Joint ALMA Observatory is operated by ESO, AUI/NRAO, and NAOJ. This work has been supported by National Key R\&D Program of China (No. 2022YFA1603101), and by NSFC through the grants No. 12033005, No. 12073061, No. 12122307, and No. 12103045. S.-L. Qin thanks the Xinjiang Uygur Autonomous Region of China for their support through the Tianchi Program. MYT acknowledges the support by NSFC through grants No.12203011, and Yunnan provincial Department of Science and Technology through grant No.202101BA070001-261. T. Zhang thanks the student's exchange program of the Collaborative Research Centre 956, funded by the Deutsche Forschungsgemeinschaft (DFG).

\section*{Data Availability}
The data underlying this article are available in the ALMA archive.

\bibliographystyle{mnras}
\bibliography{example} 

\appendix

\section{Spectra of line-rich cores}\label{sec:solc}
The spectra and optical depths of the three molecules detected in the 19 line-rich cores are shown in Figure \ref{fig1} and Figure \ref{fig8}.

\begin{figure*}
	\includegraphics[width=\linewidth]{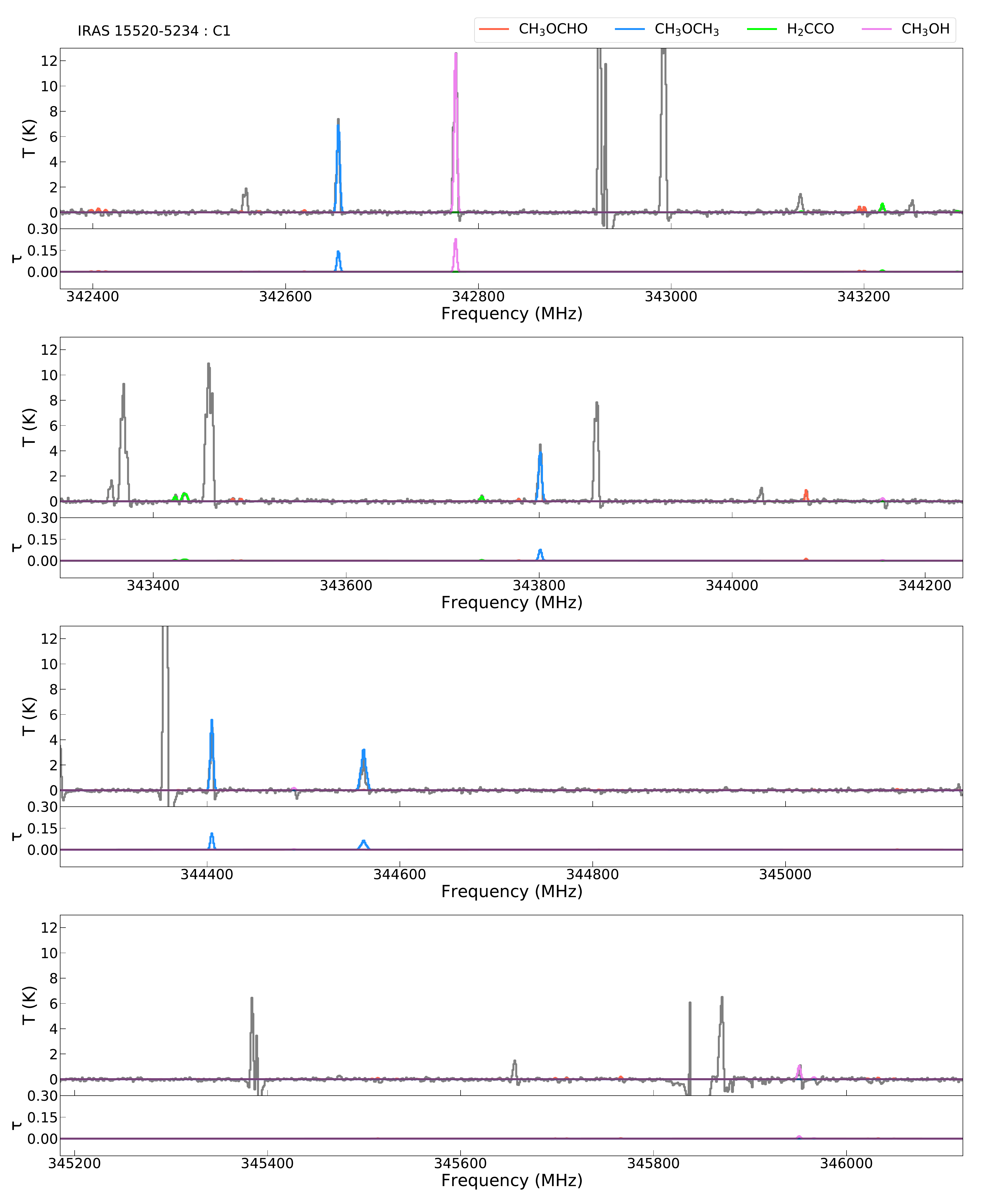}
	\caption{Same as Figure \ref{fig1}, but for the other 18 line-rich cores.} \label{fig8}
\end{figure*} 

\addtocounter{figure}{-1}
\begin{figure*}
	\includegraphics[width=\linewidth]{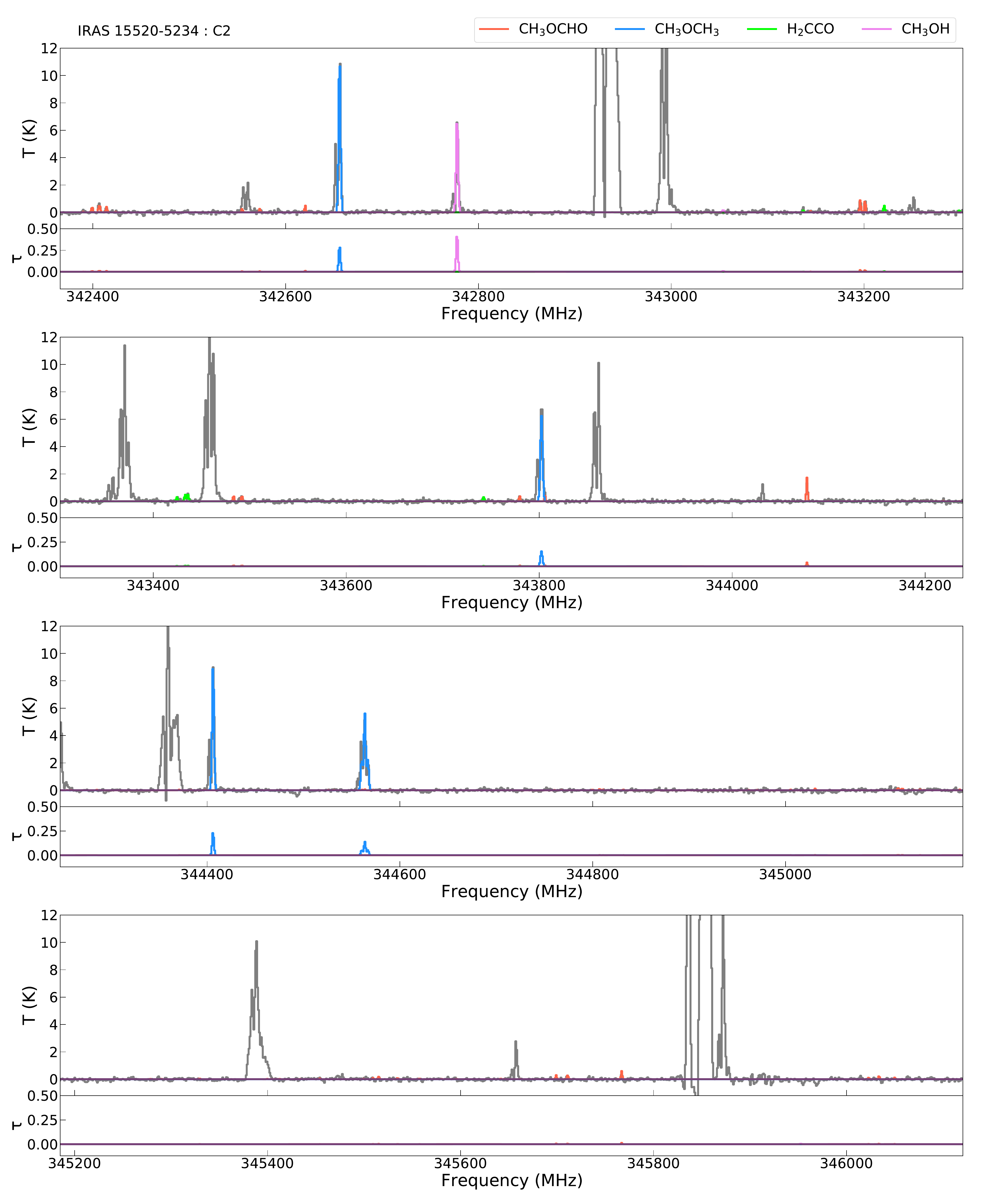}
	\caption{\it -- continued}
\end{figure*}

\addtocounter{figure}{-1}
\begin{figure*}
	\includegraphics[width=\linewidth]{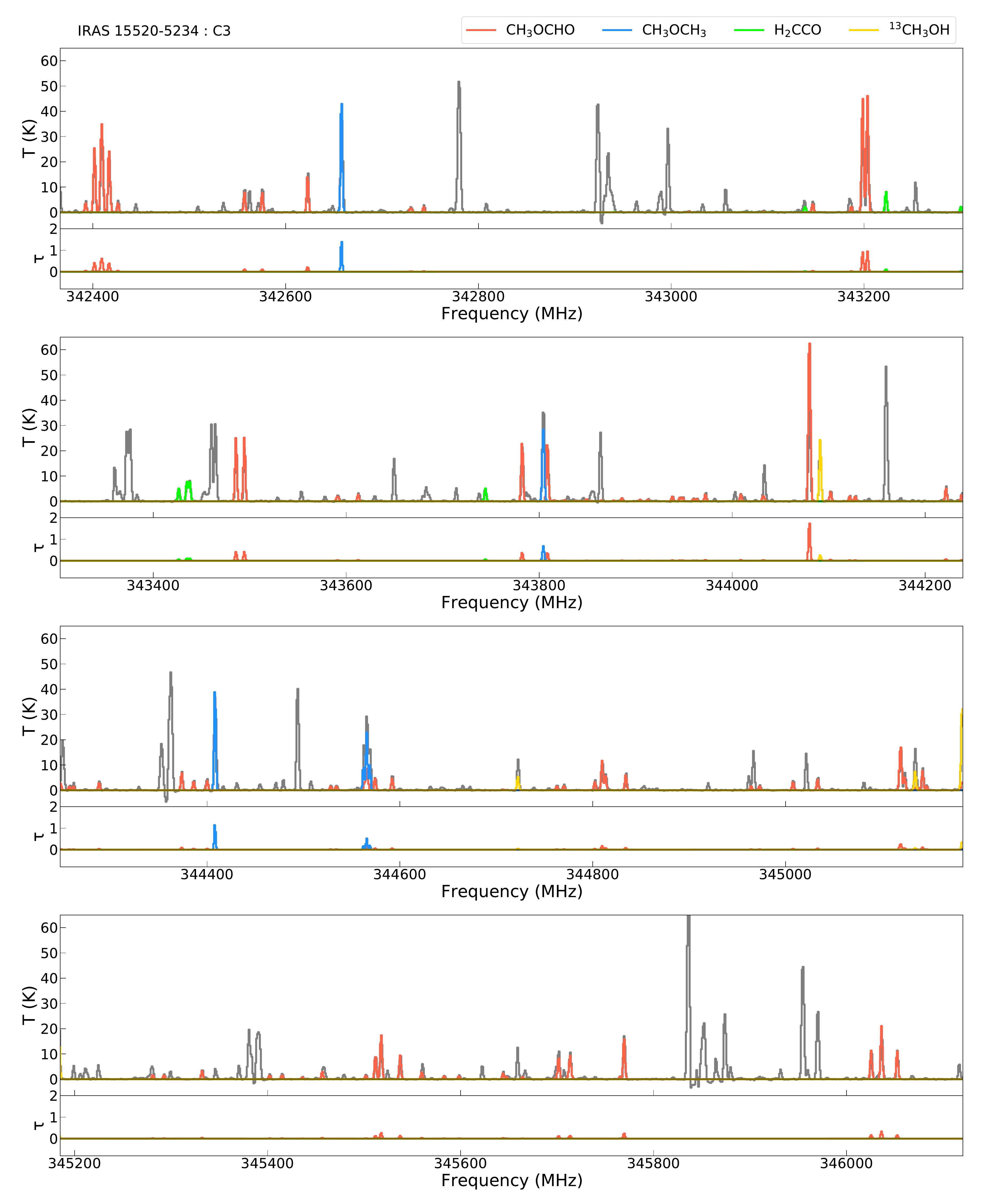}
	\caption{\it -- continued}
\end{figure*}

\addtocounter{figure}{-1}
\begin{figure*}
	\includegraphics[width=\linewidth]{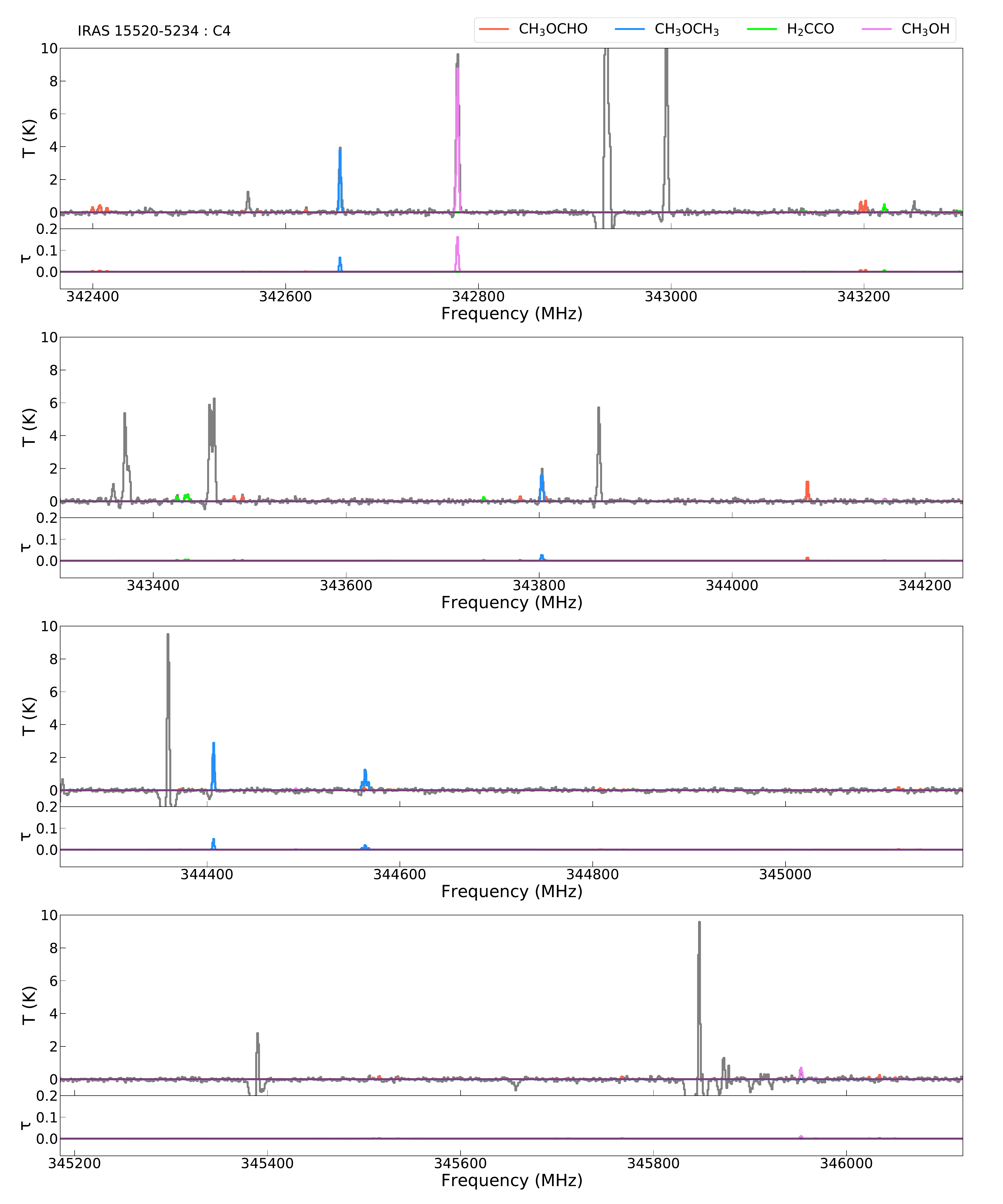}
	\caption{\it -- continued}
\end{figure*}

\addtocounter{figure}{-1}
\begin{figure*}
	\includegraphics[width=\linewidth]{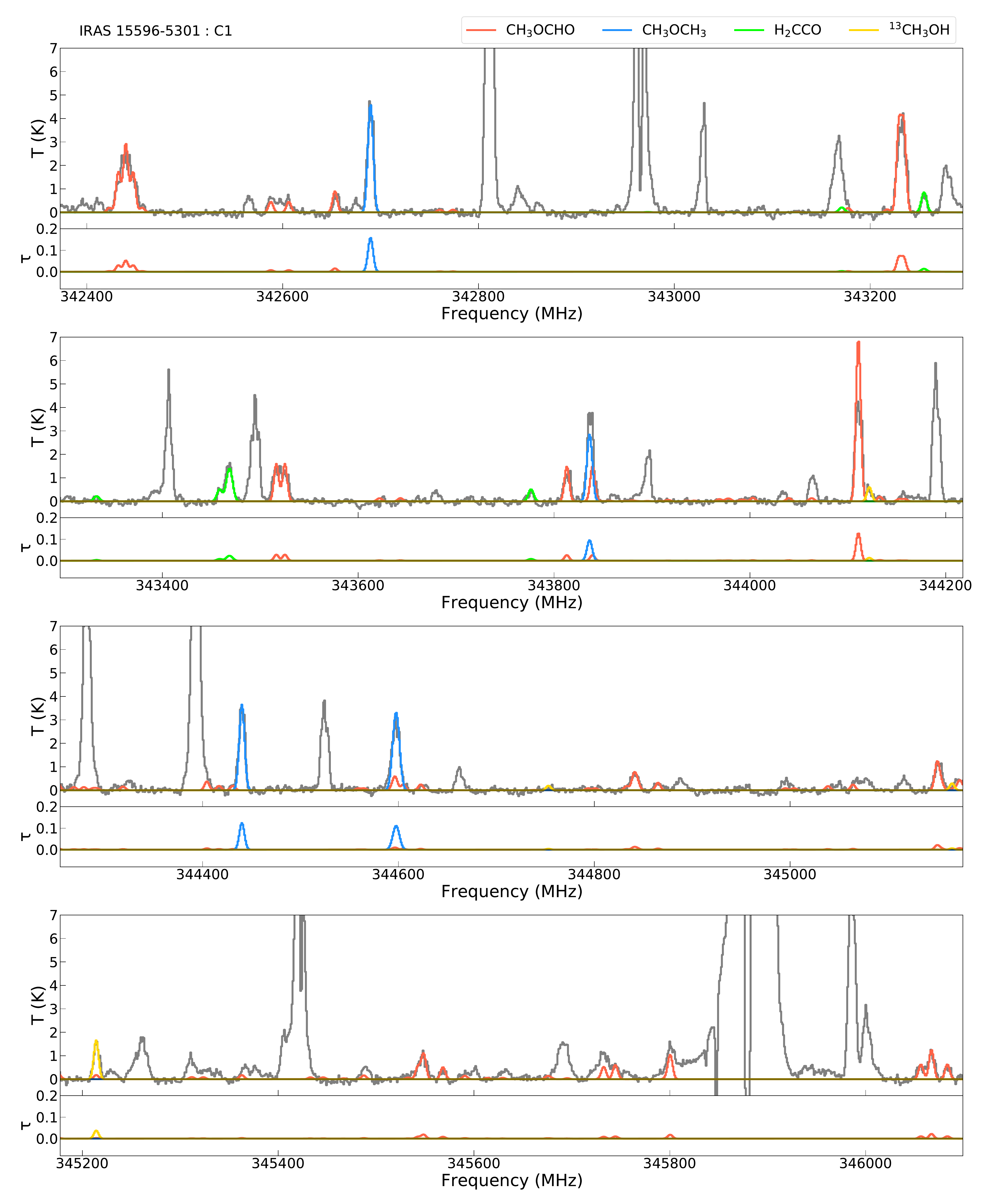}
	\caption{\it -- continued}
\end{figure*}

\addtocounter{figure}{-1}
\begin{figure*}
	\includegraphics[width=\linewidth]{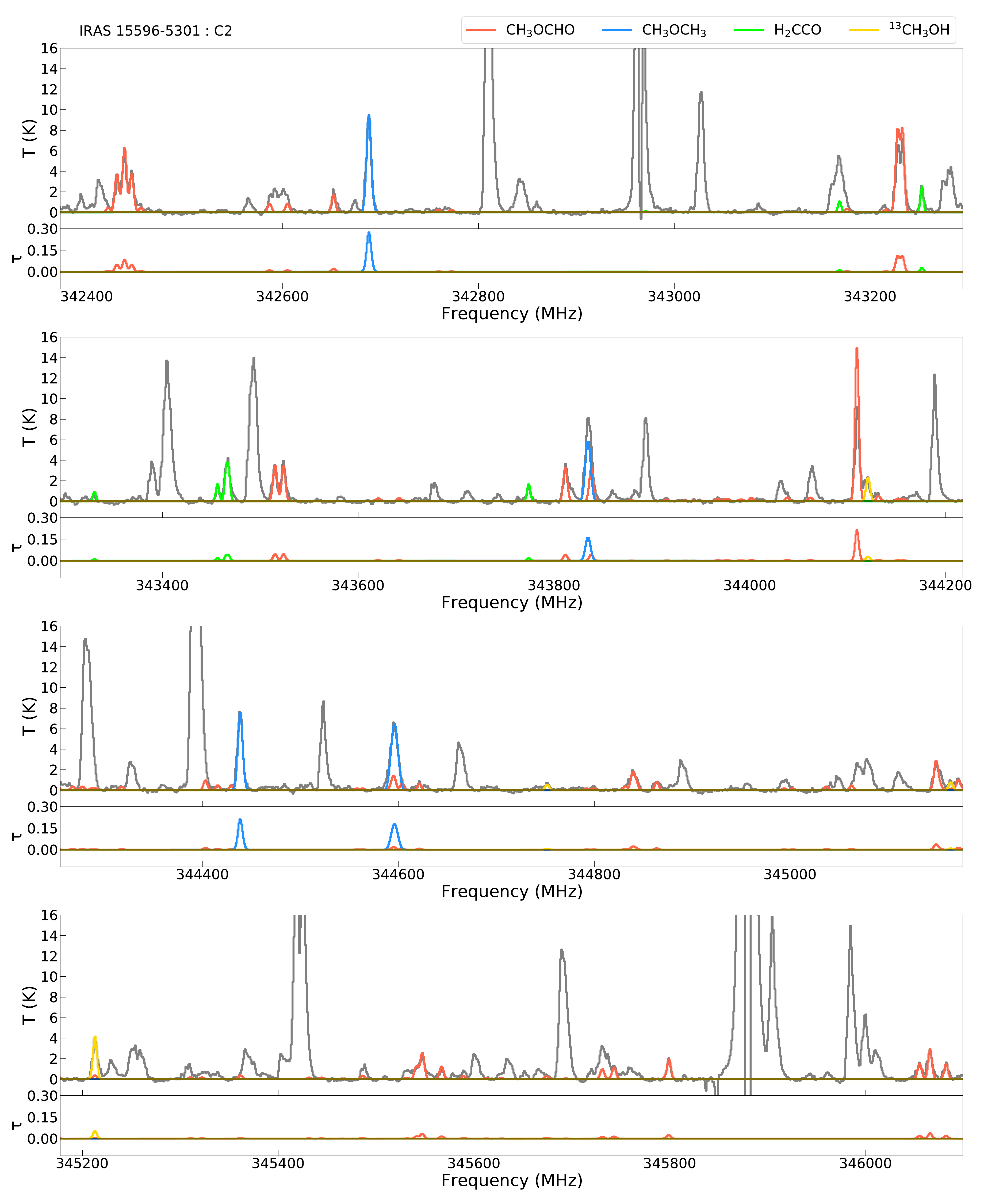}
	\caption{\it -- continued}
\end{figure*}

\addtocounter{figure}{-1}
\begin{figure*}
	\includegraphics[width=\linewidth]{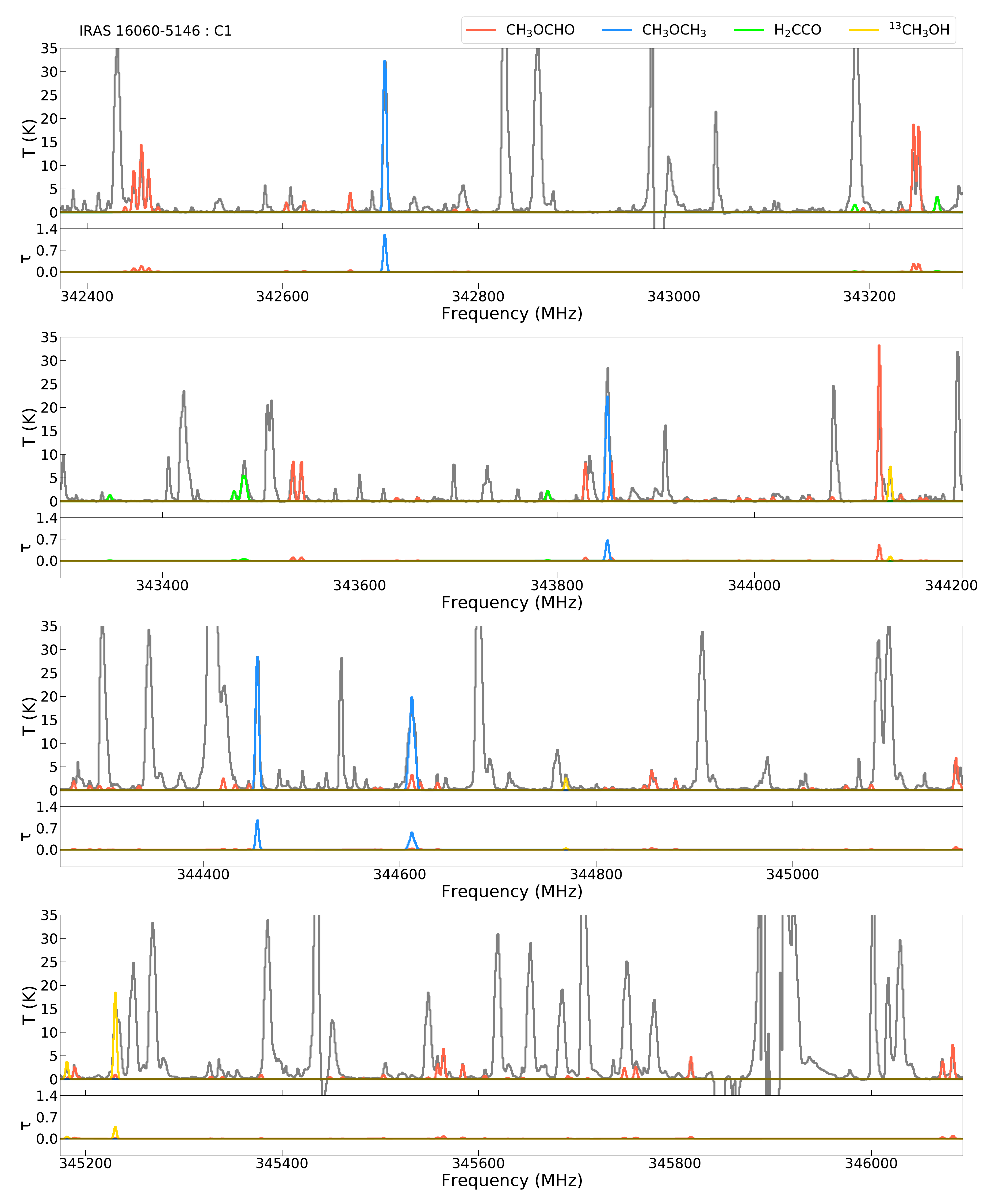}
	\caption{\it -- continued}
\end{figure*}

\addtocounter{figure}{-1}
\begin{figure*}
	\includegraphics[width=\linewidth]{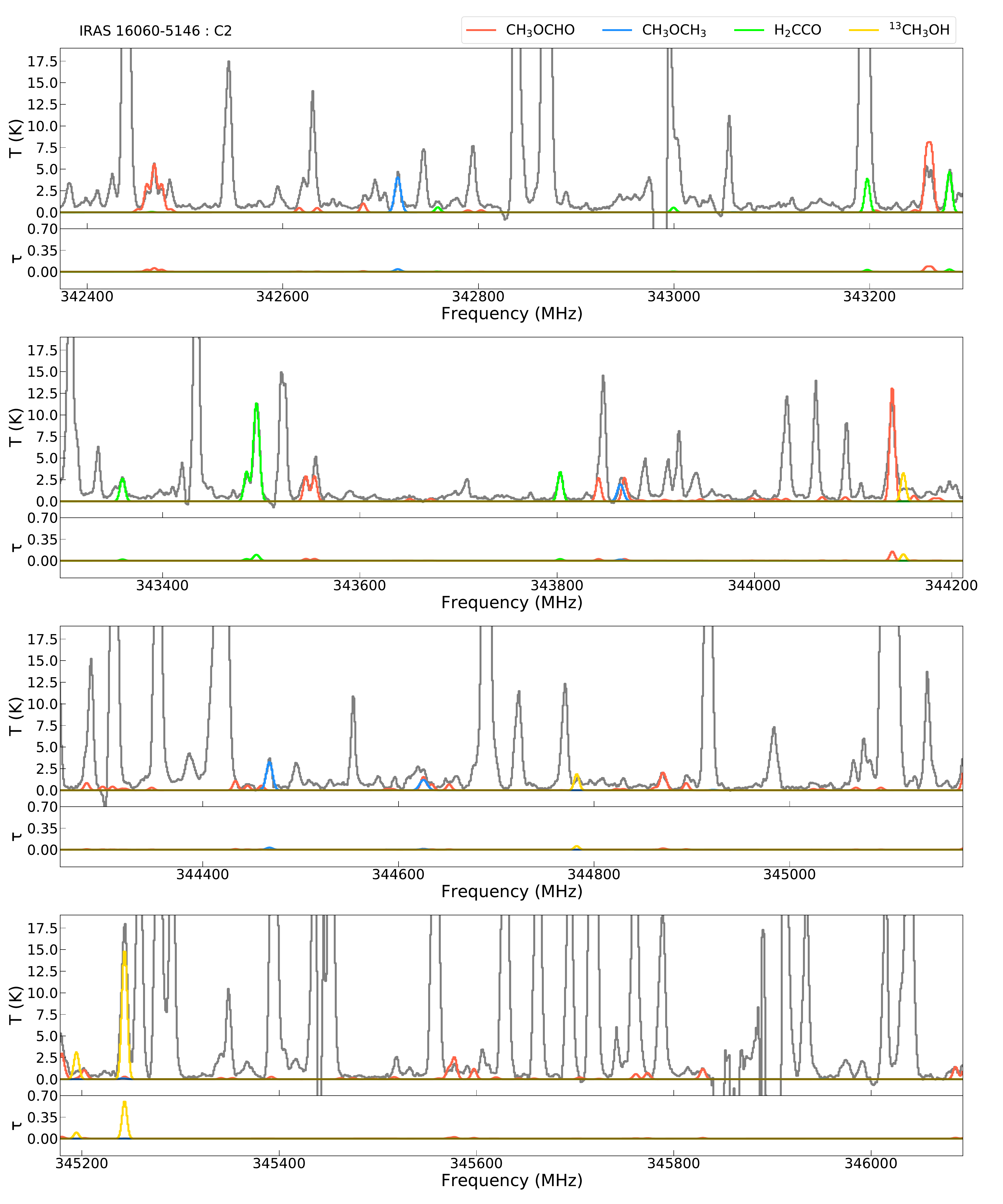}
	\caption{\it -- continued}
\end{figure*}

\addtocounter{figure}{-1}
\begin{figure*}
	\includegraphics[width=\linewidth]{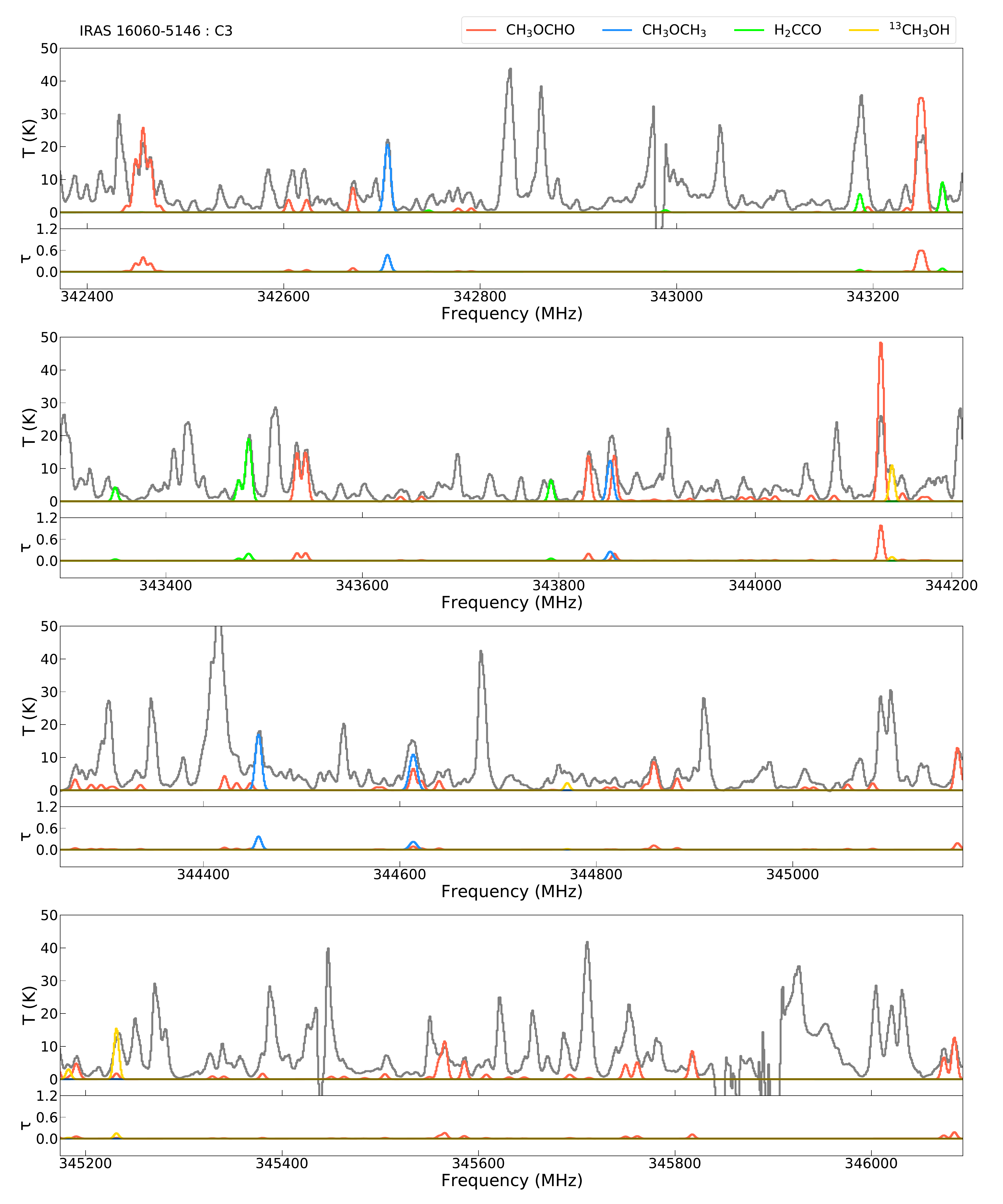}
	\caption{\it -- continued}
\end{figure*}

\addtocounter{figure}{-1}
\begin{figure*}
	\includegraphics[width=\linewidth]{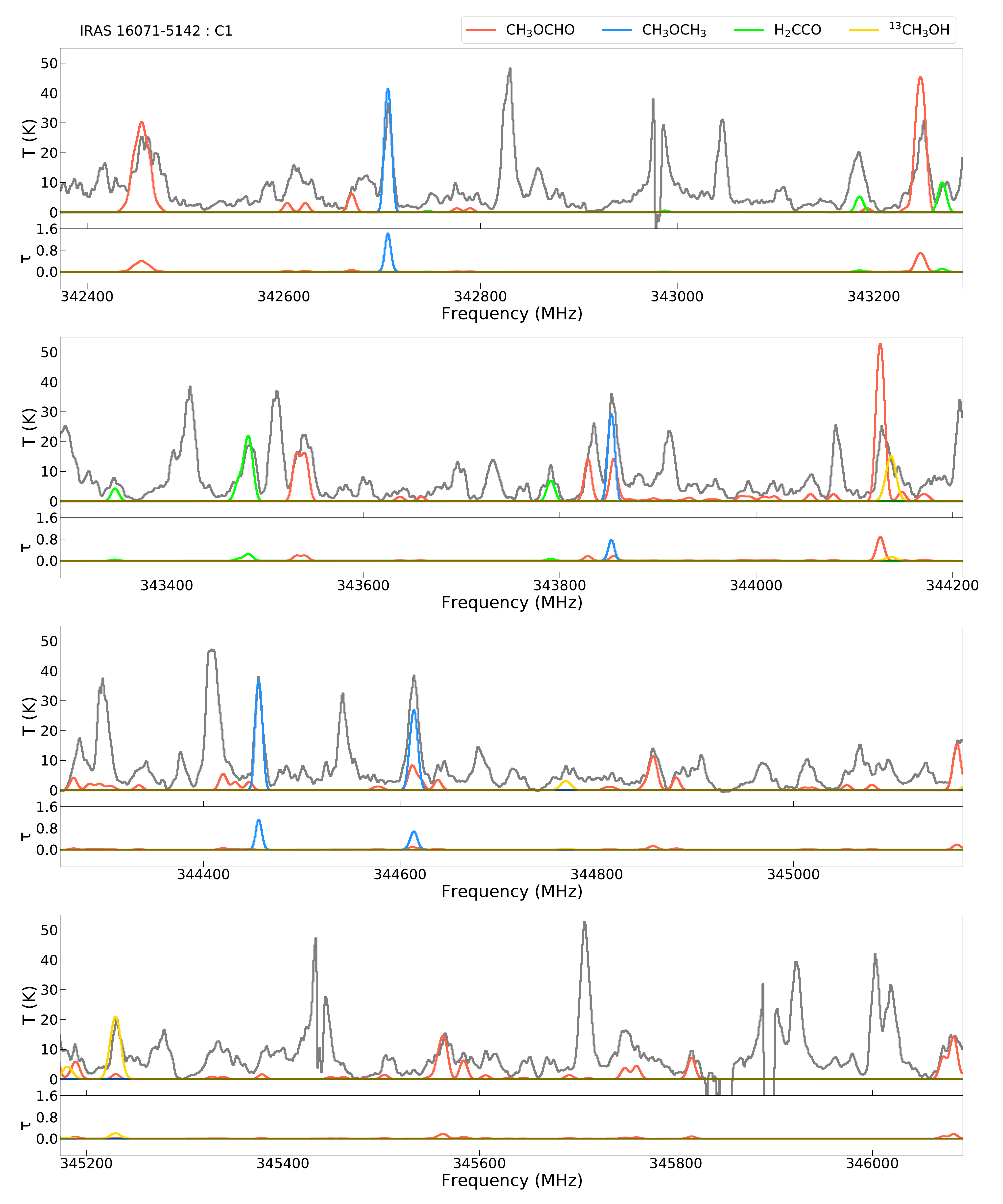}
	\caption{\it -- continued}
\end{figure*}

\addtocounter{figure}{-1}
\begin{figure*}
	\includegraphics[width=\linewidth]{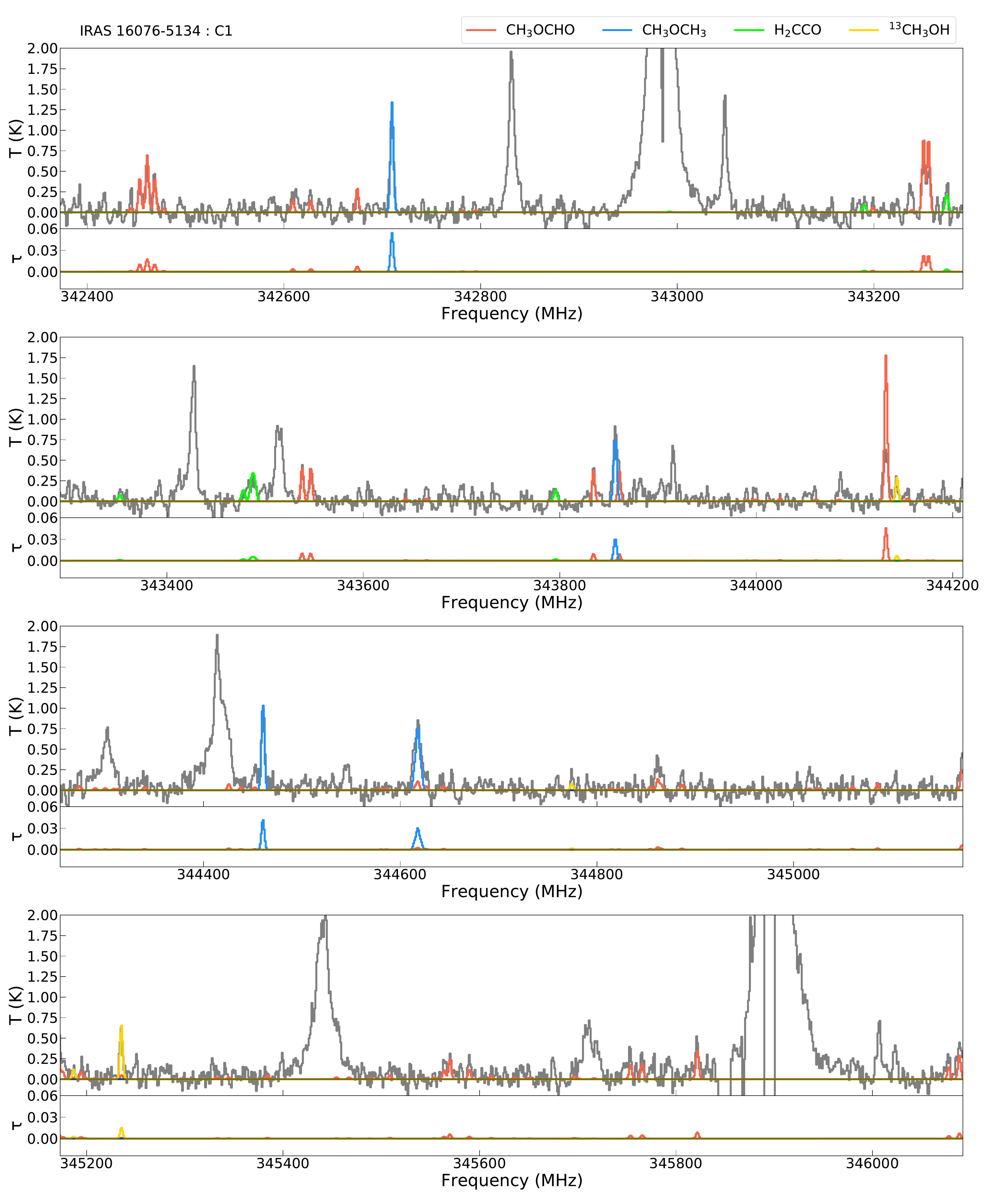}
	\caption{\it -- continued}
\end{figure*}

\addtocounter{figure}{-1}
\begin{figure*}
	\includegraphics[width=\linewidth]{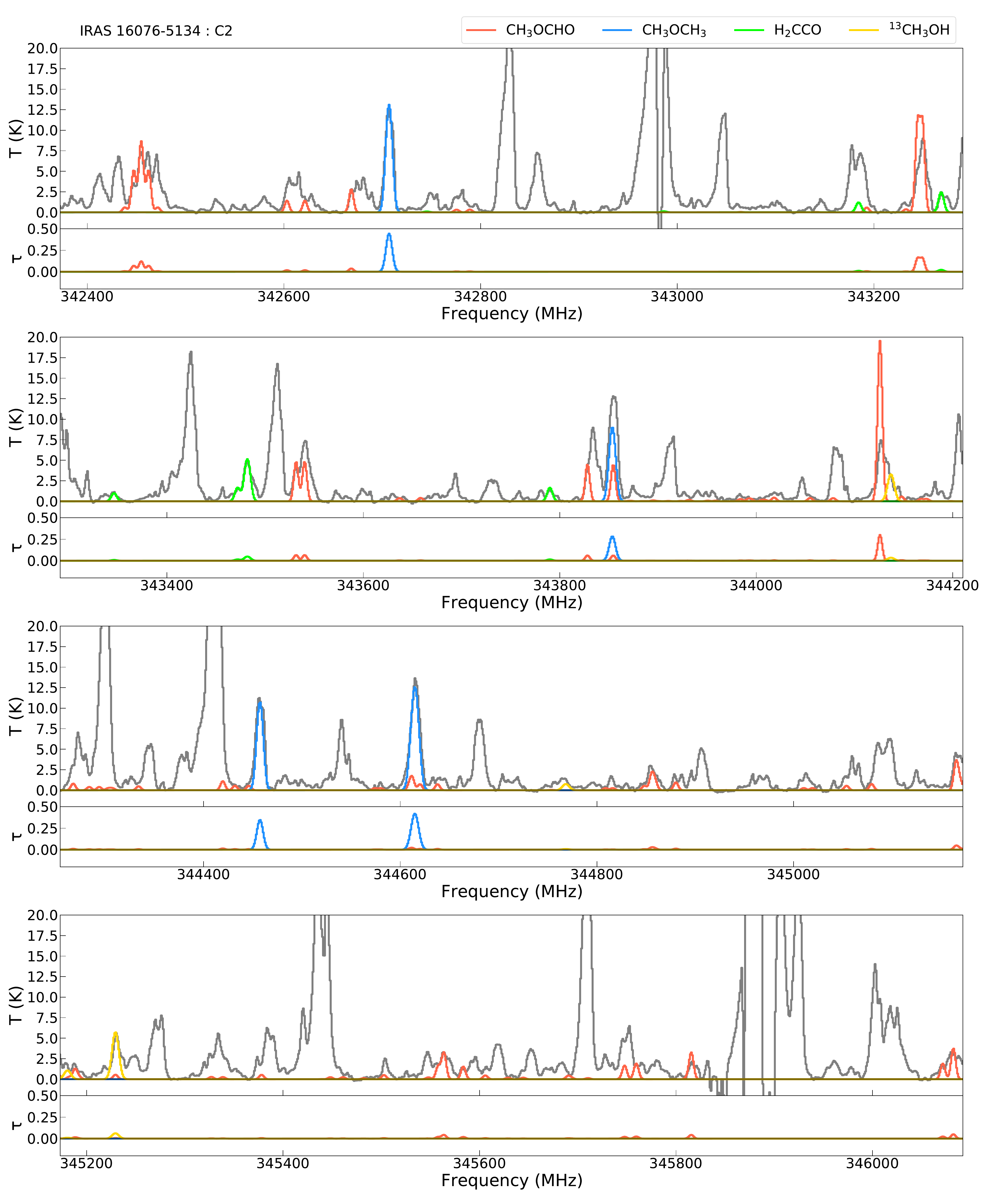}
	\caption{\it -- continued}
\end{figure*}

\addtocounter{figure}{-1}
\begin{figure*}
	\includegraphics[width=\linewidth]{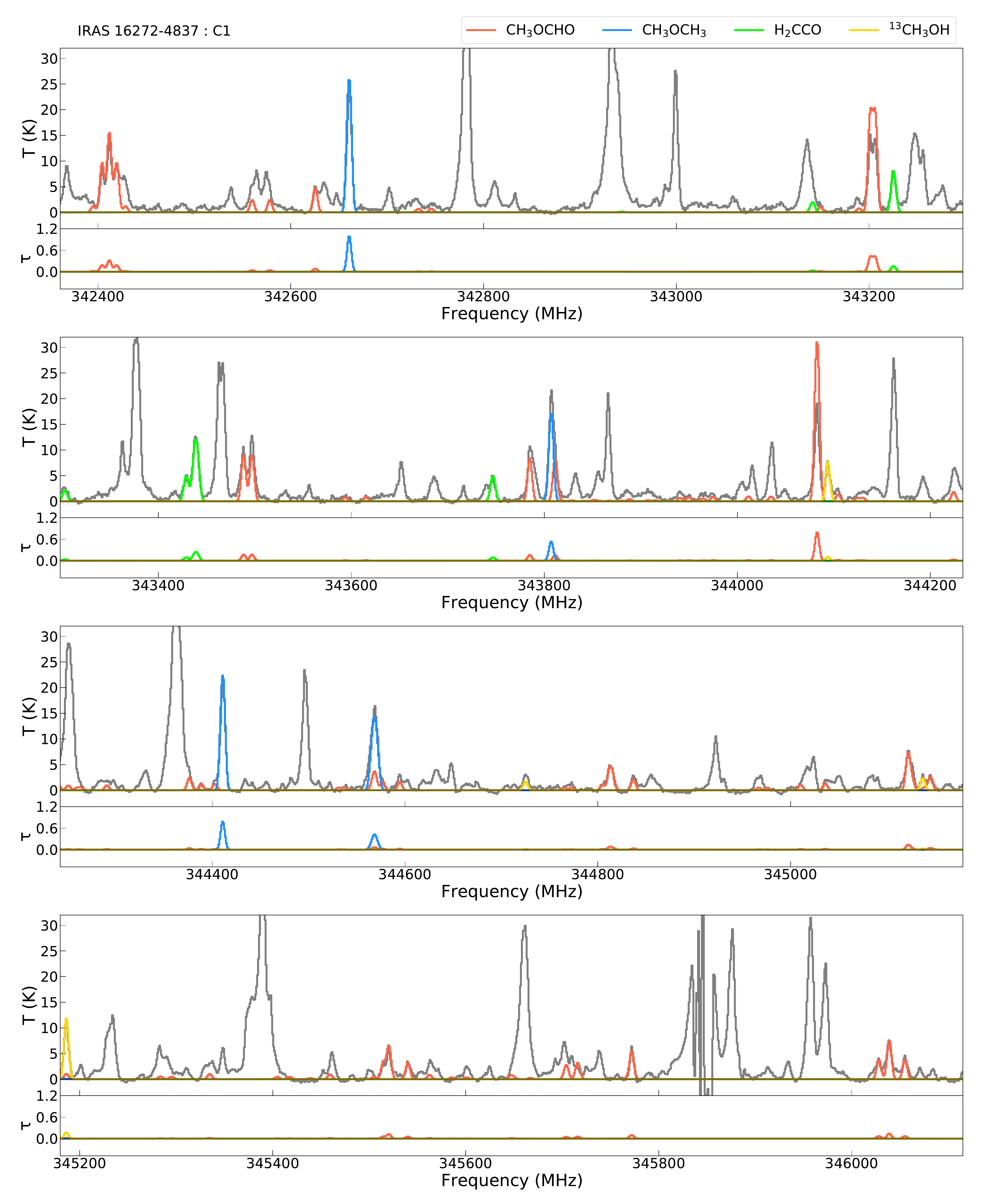}
	\caption{\it -- continued}
\end{figure*}

\addtocounter{figure}{-1}
\begin{figure*}
	\includegraphics[width=\linewidth]{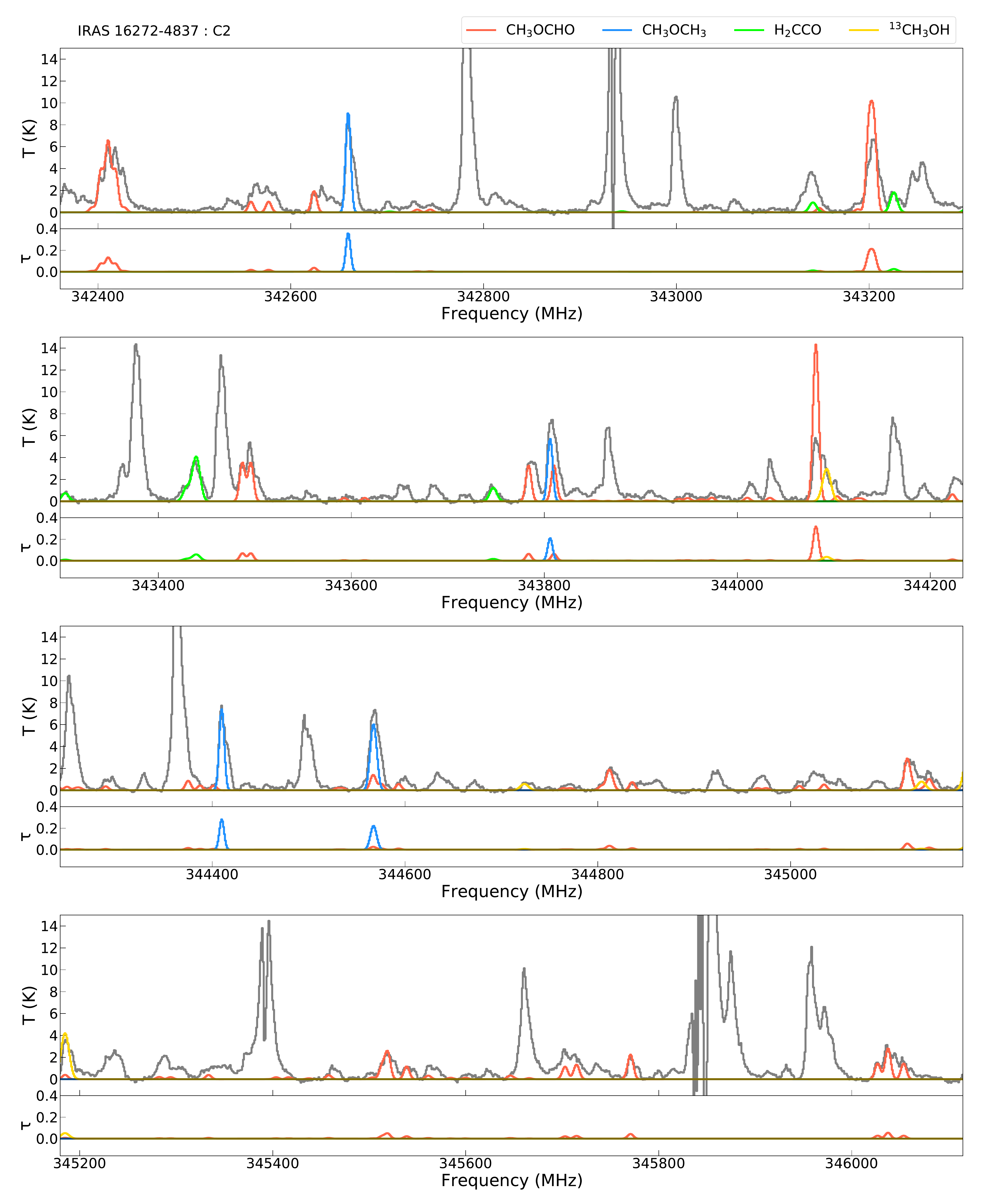}
	\caption{\it -- continued}
\end{figure*}

\addtocounter{figure}{-1}
\begin{figure*}
	\includegraphics[width=\linewidth]{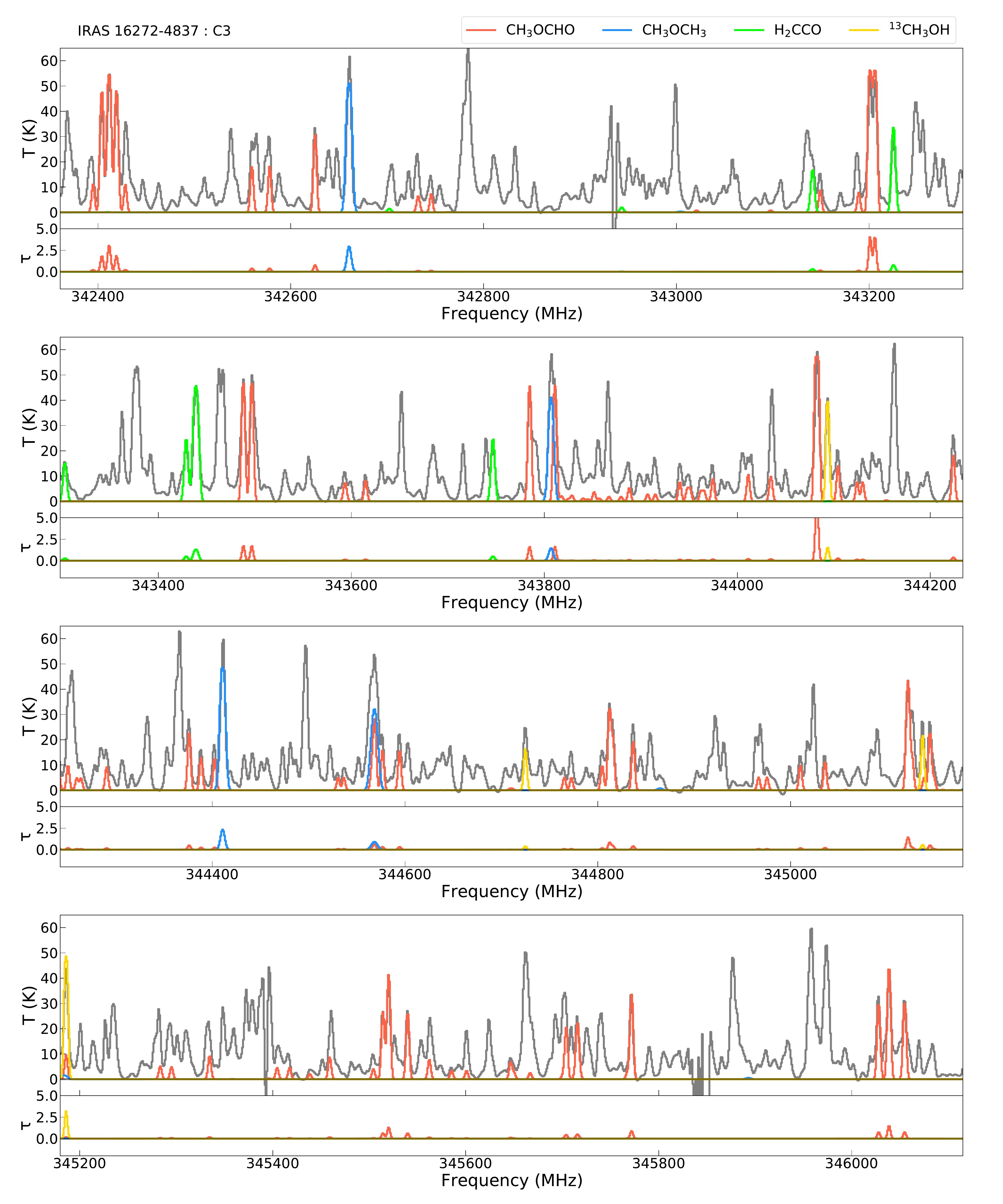}
	\caption{\it -- continued}
\end{figure*}

\addtocounter{figure}{-1}
\begin{figure*}
	\includegraphics[width=\linewidth]{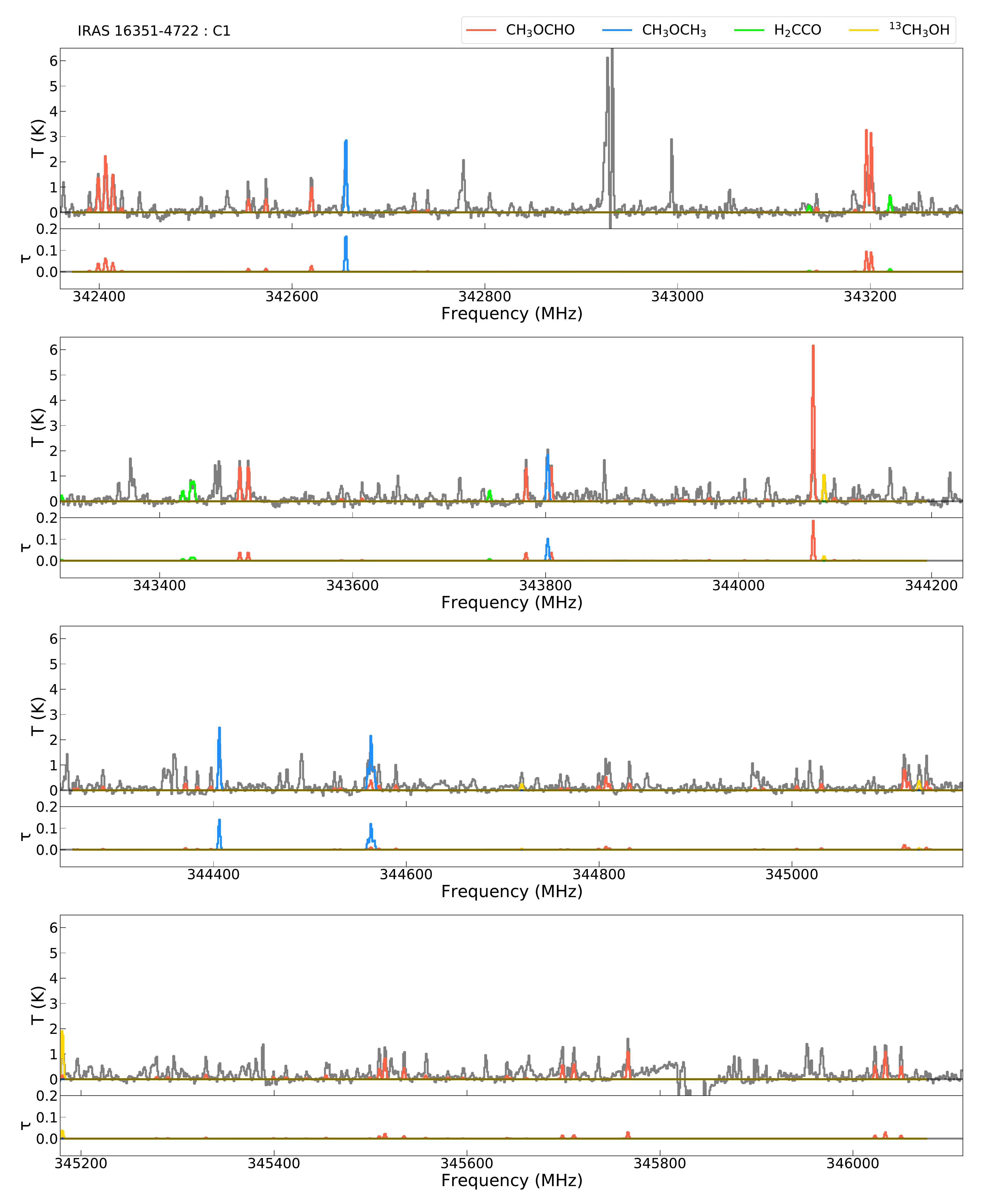}
	\caption{\it -- continued}
\end{figure*}

\addtocounter{figure}{-1}
\begin{figure*}
	\includegraphics[width=\linewidth]{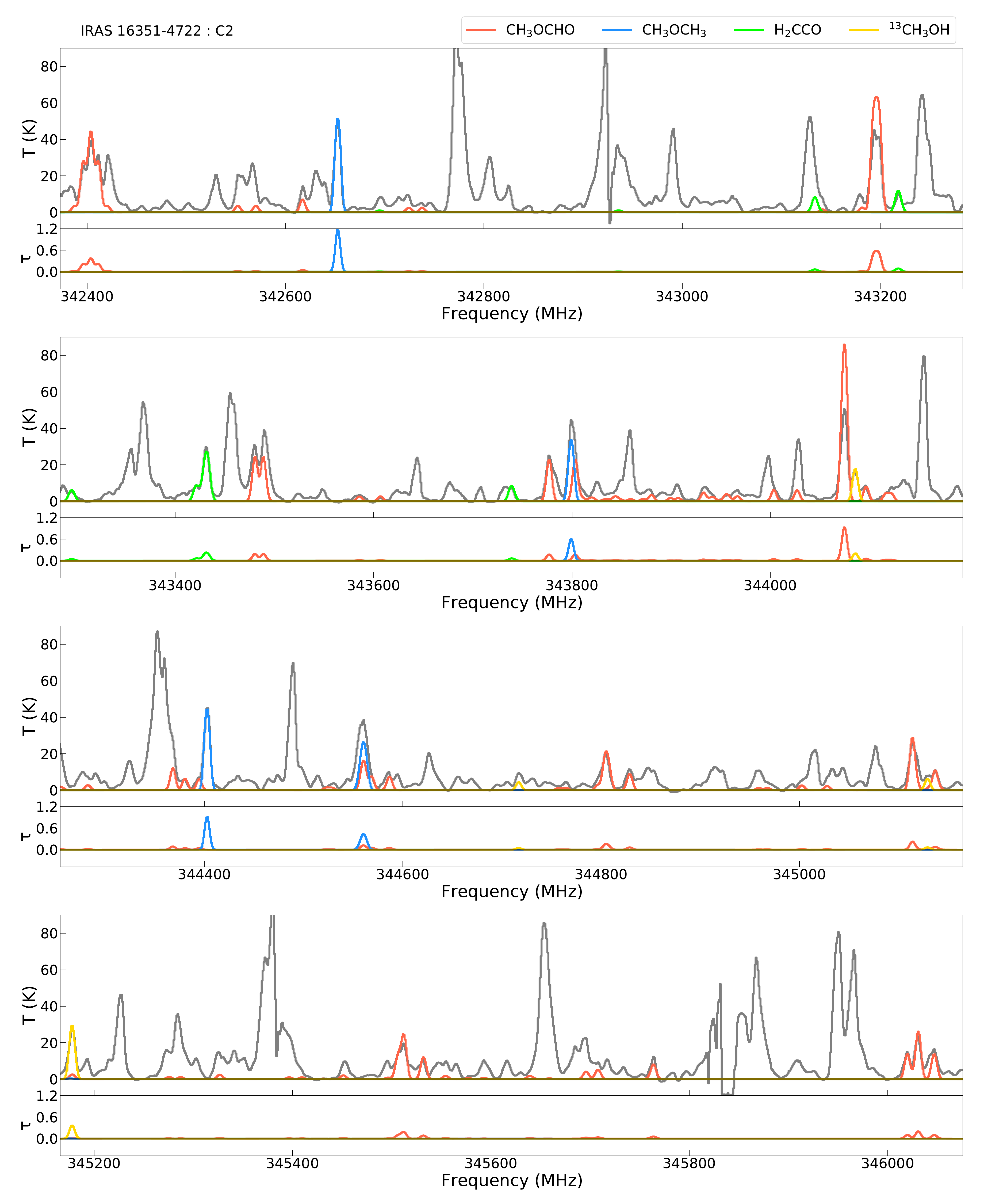}
	\caption{\it -- continued}
\end{figure*}

\addtocounter{figure}{-1}
\begin{figure*}
	\includegraphics[width=\linewidth]{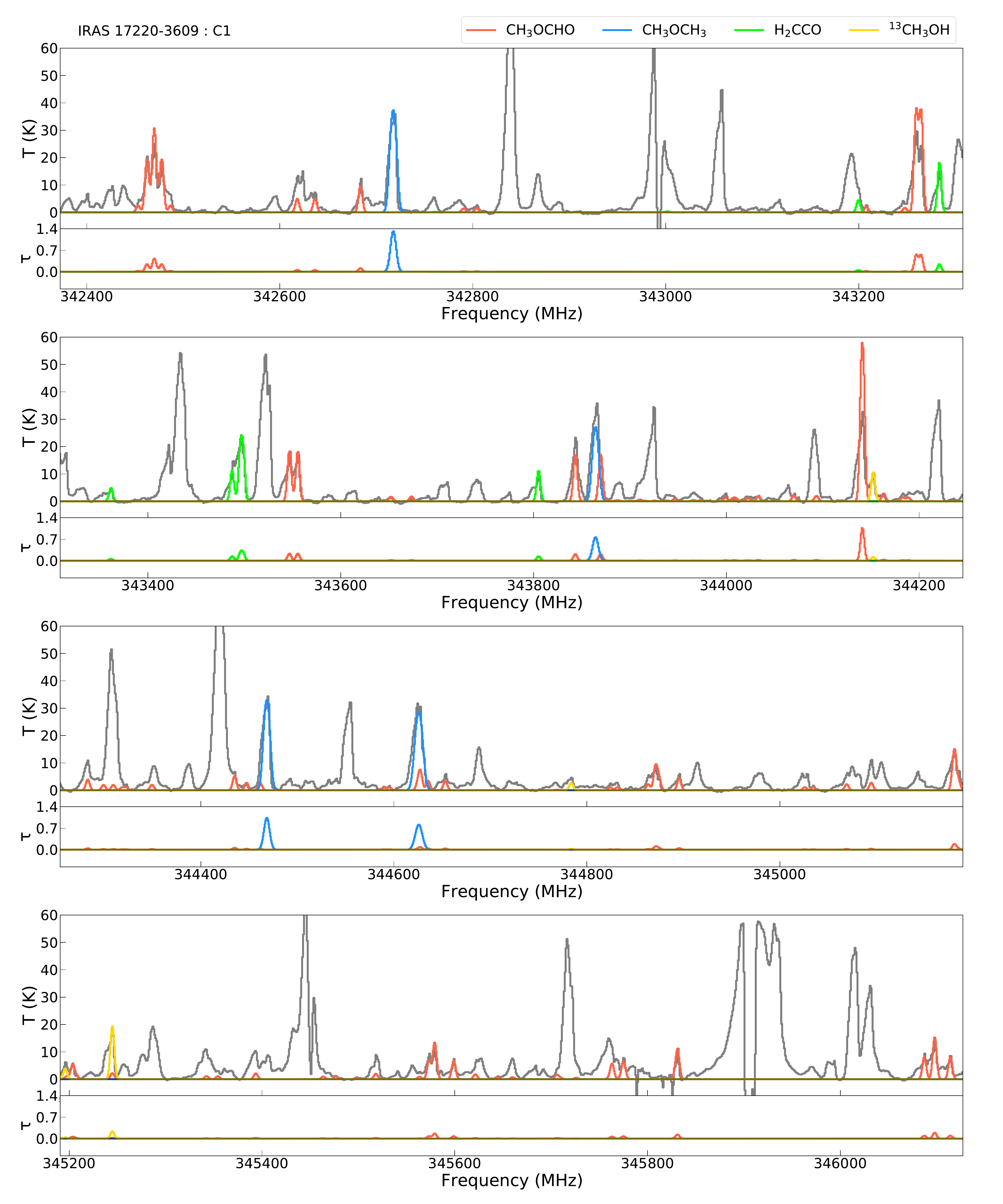}
	\caption{\it -- continued}
\end{figure*}

\section{Molecular Transitions}\label{sec:mt}
The rotational transitions of the three molecules detected in 19 dense cores are shown in Table \ref{tab4}. The molecular parameters are taken from the JPL catalogue for \MF\ lines and the CDMS catalogue for \DE\ and \K\ lines. Due to internal rotation, the \MF\ rotational levels split into two substates A and E, while the \DE\ rotational levels split into four substates AA, EE, EA, and AE.

\begin{table}
	\centering
	\caption{The Detected Transitions of \MF, \DE, and \K\ with Signal Above 3$\sigma$ Noise Level.}
	\label{tab4}
	\begin{tabular}{lccc}
		\hline\hline
		Frequency & \multirow{2}{*}{Transition} & S$\mu$$^{2}$ & E$_{\rm up}$\\
		(MHz) && (D$^{2}$) & (K)\\
		\hline
		\multicolumn{4}{c}{\MF\ (v=0)}                 									\\
		\cline{1-4}									
		342342.185 	&	30(2,28)$-$29(3,27) E	&	10.00 	&	269.50 	\\
		342350.119 	&	30(2,28)$-$29(3,27) A	&	10.00 	&	269.49 	\\
		342351.420 	&	30(3,28)$-$29(3,27) E	&	78.04 	&	269.50 	\\
		342358.225 	&	30(2,28)$-$29(2,27) E	&	78.03 	&	269.50 	\\
		342359.508 	&	30(3,28)$-$29(3,27) A	&	78.03 	&	269.49 	\\
		342366.296 	&	30(2,28)$-$29(2,27) A	&	78.03 	&	269.49 	\\
		342367.680 	&	30(3,28)$-$29(2,27) E	&	10.00 	&	269.50 	\\
		342375.658 	&	30(3,28)$-$29(2,27) A	&	10.00 	&	269.49 	\\
		342506.986 	&	11(8,4)$-$10(7,4) E	&	3.46 	&	81.40 	\\
		342525.299 	&	11(8,3)$-$10(7,3) E	&	3.46 	&	81.42 	\\
		342572.422 	&	11(8,4)$-$10(7,3) A	&	3.46 	&	81.41 	\\
		342572.422 	&	11(8,3)$-$10(7,4) A	&	3.46 	&	81.41 	\\
		342678.725 	&	27(13,15)$-$27(12,16) E	&	5.56 	&	335.27 	\\
		342680.167 	&	27(13,14)$-$27(12,15) E	&	5.56 	&	335.28 	\\
		342692.876 	&	27(13,14)$-$27(12,15) A	&	5.56 	&	335.28 	\\
		342692.888 	&	27(13,15)$-$27(12,16) A	&	5.56 	&	335.28 	\\
		343096.392 	&	17(5,12)$-$16(4,13) E	&	1.87 	&	107.82 	\\
		343136.353 	&	26(13,13)$-$26(12,14) E	&	5.23 	&	319.30 	\\
		343136.356 	&	26(13,14)$-$26(12,15) E	&	5.23 	&	319.29 	\\
		343147.898 	&	31(1,30)$-$30(2,29) E	&	11.85 	&	273.44 	\\
		343148.047 	&	31(2,30)$-$30(2,29) E	&	81.50 	&	273.44 	\\
		343148.169 	&	31(1,30)$-$30(1,29) E	&	81.50 	&	273.44 	\\
		343148.318 	&	31(2,30)$-$30(1,29) E	&	11.85 	&	273.44 	\\
		343149.303 	&	17(5,12)$-$16(4,13) A	&	1.87 	&	107.81 	\\
		343152.958 	&	31(1,30)$-$30(2,29) A	&	11.89 	&	273.43 	\\
		343153.106 	&	31(2,30)$-$30(2,29) A	&	81.46 	&	273.43 	\\
		343153.227 	&	31(1,30)$-$30(1,29) A	&	81.46 	&	273.43 	\\
		343153.376 	&	31(2,30)$-$30(1,29) A	&	11.89 	&	273.43 	\\
		343153.762 	&	26(13,13)$-$26(12,14) A	&	5.23 	&	319.30 	\\
		343153.770 	&	26(13,14)$-$26(12,15) A	&	5.23 	&	319.30 	\\
		343435.260 	&	28(4,24)$-$27(4,23) E	&	71.57 	&	257.08 	\\
		343443.944 	&	28(4,24)$-$27(4,23) A	&	71.58 	&	257.08 	\\
		343539.815 	&	25(13,12)$-$25(12,13) E	&	4.90 	&	303.92 	\\
		343541.355 	&	25(13,13)$-$25(12,14) E	&	4.90 	&	303.91 	\\
		343561.850 	&	25(13,12)$-$25(12,13) A	&	4.90 	&	303.92 	\\
		343561.883 	&	25(13,13)$-$25(12,14) A	&	4.90 	&	303.92 	\\
		343731.783 	&	27(7,20)$-$26(7,19) E	&	67.17 	&	258.47 	\\
		343758.010 	&	27(7,20)$-$26(7,19) A	&	67.19 	&	258.48 	\\
		343798.647 	&	28(23,5)$-$27(23,4) A	&	24.44 	&	589.86 	\\
		343798.647 	&	28(23,6)$-$27(23,5) A	&	24.44 	&	589.86 	\\
		343814.031 	&	28(23,5)$-$27(23,4) E	&	24.44 	&	589.86 	\\
		343826.593 	&	28(23,6)$-$27(23,5) E	&	24.44 	&	589.85 	\\
		343835.114 	&	28(22,7)$-$27(22,6) A	&	28.73 	&	560.10 	\\
		343835.114 	&	28(22,6)$-$27(22,5) A	&	28.73 	&	560.10 	\\
		343854.153 	&	28(22,6)$-$27(22,5) E	&	28.74 	&	560.10 	\\
		343862.096 	&	28(22,7)$-$27(22,6) E	&	28.73 	&	560.09 	\\
		343887.467 	&	28(21,8)$-$27(21,7) A	&	32.84 	&	531.66 	\\
		343887.483 	&	28(21,7)$-$27(21,6) A	&	32.84 	&	531.66 	\\
		343895.152 	&	24(13,11)$-$24(12,12) E	&	4.57 	&	289.13 	\\
		343898.142 	&	24(13,12)$-$24(12,13) E	&	4.57 	&	289.13 	\\
		343909.244 	&	28(21,7)$-$27(21,6) E	&	32.83 	&	531.65 	\\
		343912.685 	&	28(21,8)$-$27(21,7) E	&	32.84 	&	531.65 	\\
		343921.695 	&	24(13,11)$-$24(12,12) A	&	4.57 	&	289.14 	\\
		343921.695 	&	24(13,12)$-$24(12,13) A	&	4.57 	&	289.14 	\\
		343958.362 	&	28(20,9)$-$27(20,8) A	&	36.75 	&	504.53 	\\
		343958.362 	&	28(20,8)$-$27(20,7) A	&	36.75 	&	504.53 	\\
		343981.148 	&	28(20,9)$-$27(20,8) E	&	36.74 	&	504.52 	\\
		343982.450 	&	28(20,8)$-$27(20,7) E	&	36.74 	&	504.52 	\\
		344029.259 	&	32(0,32)$-$31(1,31) E	&	10.53 	&	276.10 	\\
		344029.260 	&	32(1,32)$-$31(1,31) E	&	88.18 	&	276.10 	\\
		\hline
	\end{tabular}
\end{table}

\addtocounter{table}{-1}
\begin{table}
	\centering
	\caption{\it -- continued}
	\begin{tabular}{lccc}
		\hline\hline
		Frequency & \multirow{2}{*}{Transition} & S$\mu$$^{2}$ & E$_{\rm up}$\\
		(MHz) && (D$^{2}$) & (K)\\
		\hline
		\multicolumn{4}{c}{\MF\ (v=0)}                  									\\
		\cline{1-4}										
		344029.261 	&	32(0,32)$-$31(0,31) E	&	88.18 	&	276.10 	\\
		344029.262 	&	32(1,32)$-$31(0,31) E	&	10.53 	&	276.10 	\\
		344029.645 	&	32(0,32)$-$31(1,31) A	&	13.73 	&	276.08 	\\
		344029.645 	&	32(1,32)$-$31(1,31) A	&	84.98 	&	276.08 	\\
		344029.646 	&	32(0,32)$-$31(0,31) A	&	84.98 	&	276.08 	\\
		344029.647 	&	32(1,32)$-$31(0,31) A	&	13.73 	&	276.08 	\\
		344051.371 	&	28(19,10)$-$27(19,9) A	&	40.46 	&	478.72 	\\
		344051.371 	&	28(19,9)$-$27(19,8) A	&	40.46 	&	478.72 	\\
		344071.196 	&	28(19,9)$-$27(19,8) E	&	40.45 	&	478.72 	\\
		344076.934 	&	28(19,10)$-$27(19,9) E	&	40.45 	&	478.72 	\\
		344170.930 	&	28(18,10)$-$27(18,9) A	&	43.97 	&	454.24 	\\
		344170.930 	&	28(18,11)$-$27(18,10) A	&	43.97 	&	454.24 	\\
		344187.246 	&	28(18,10)$-$27(18,9) E	&	43.98 	&	454.24 	\\
		344197.339 	&	28(18,11)$-$27(18,10) E	&	43.97 	&	454.24 	\\
		344206.436 	&	23(13,10)$-$23(12,11) E	&	4.23 	&	274.95 	\\
		344210.812 	&	23(13,11)$-$23(12,12) E	&	4.23 	&	274.94 	\\
		344237.391 	&	23(13,10)$-$23(12,11) A	&	4.23 	&	274.95 	\\
		344237.414 	&	23(13,11)$-$23(12,12) A	&	4.23 	&	274.95 	\\
		344322.992 	&	28(17,12)$-$27(17,11) A	&	47.30 	&	431.09 	\\
		344322.992 	&	28(17,11)$-$27(17,10) A	&	47.30 	&	431.09 	\\
		344335.357 	&	28(17,11)$-$27(17,10) E	&	47.30 	&	431.08 	\\
		344349.515 	&	28(17,12)$-$27(17,11) E	&	47.30 	&	431.08 	\\
		344477.576 	&	22(13,9)$-$22(12,10) E	&	3.89 	&	261.36 	\\
		344483.461 	&	22(13,10)$-$22(12,11) E	&	3.89 	&	261.35 	\\
		344512.739 	&	22(13,9)$-$22(12,10) A	&	3.89 	&	261.37 	\\
		344512.747 	&	22(13,10)$-$22(12,11) A	&	3.89 	&	261.37 	\\
		344515.454 	&	28(16,13)$-$27(16,12) A	&	50.43 	&	409.27 	\\
		344515.454 	&	28(16,12)$-$27(16,11) A	&	50.43 	&	409.27 	\\
		344523.525 	&	28(16,12)$-$27(16,11) E	&	50.43 	&	409.26 	\\
		344541.314 	&	28(16,13)$-$27(16,12) E	&	50.43 	&	409.26 	\\
		344712.139 	&	21(13,8)$-$21(12,9) E	&	3.55 	&	248.37 	\\
		344719.465 	&	21(13,9)$-$21(12,10) E	&	3.55 	&	248.36 	\\
		344751.514 	&	21(13,9)$-$21(12,10) A	&	3.55 	&	248.37 	\\
		344751.524 	&	21(13,8)$-$21(12,9) A	&	3.55 	&	248.37 	\\
		344759.096 	&	28(15,14)$-$27(15,13) A	&	53.37 	&	388.78 	\\
		344759.096 	&	28(15,13)$-$27(15,12) A	&	53.37 	&	388.78 	\\
		344762.590 	&	28(15,13)$-$27(15,12) E	&	53.37 	&	388.78 	\\
		344783.597 	&	28(15,14)$-$27(15,13) E	&	53.38 	&	388.78 	\\
		344913.684 	&	20(13,7)$-$20(12,8) E	&	3.20 	&	235.98 	\\
		344922.458 	&	20(13,8)$-$20(12,9) E	&	3.20 	&	235.97 	\\
		344957.101 	&	20(13,8)$-$20(12,9) A	&	3.20 	&	235.98 	\\
		344957.127 	&	20(13,7)$-$20(12,8) A	&	3.20 	&	235.98 	\\
		344982.605 	&	47(6,41)$-$47(5,42) E	&	2.31 	&	104.43 	\\
		345067.795 	&	28(14,14)$-$27(14,13) E	&	56.12 	&	369.64 	\\
		345069.059 	&	28(14,15)$-$27(14,14) A	&	56.12 	&	369.64 	\\
		345069.059 	&	28(14,14)$-$27(14,13) A	&	56.12 	&	369.64 	\\
		345073.057 	&	16(6,11)$-$15(5,10) A	&	2.71 	&	104.42 	\\
		345085.394 	&	19(13,6)$-$19(12,7) E	&	2.85 	&	224.18 	\\
		345091.465 	&	28(14,15)$-$27(14,14) E	&	56.12 	&	369.64 	\\
		345095.559 	&	19(13,7)$-$19(12,8) E	&	2.85 	&	224.17 	\\
		345132.629 	&	19(13,6)$-$19(12,7) A	&	2.85 	&	224.18 	\\
		345132.655 	&	19(13,7)$-$19(12,8) A	&	2.85 	&	224.18 	\\
		345230.296 	&	18(13,5)$-$18(12,6) E	&	2.48 	&	212.97 	\\
		345241.935 	&	18(13,6)$-$18(12,7) E	&	2.48 	&	212.96 	\\
		345281.320 	&	18(13,6)$-$18(12,7) A	&	2.48 	&	212.97 	\\
		345281.320 	&	18(13,5)$-$18(12,6) A	&	2.48 	&	212.97 	\\
		345351.333 	&	17(13,4)$-$17(12,5) E	&	2.11 	&	202.36 	\\
		345364.235 	&	17(13,5)$-$17(12,6) E	&	2.11 	&	202.34 	\\
		345385.268 	&	16(6,10)$-$15(5,10) E	&	0.40 	&	104.45 	\\
		345405.869 	&	17(13,5)$-$17(12,6) A	&	2.11 	&	202.36 	\\
		\hline
	\end{tabular}
\end{table}

\addtocounter{table}{-1}
\begin{table}
	\centering
	\caption{\it -- continued}
	\begin{tabular}{lccc}
		\hline\hline
		Frequency & \multirow{2}{*}{Transition} & S$\mu$$^{2}$ & E$_{\rm up}$\\
		(MHz) && (D$^{2}$) & (K)\\
				\hline
		\multicolumn{4}{c}{\MF\ (v=0)}                 									\\
		\cline{1-4}														
		345405.869 	&	17(13,4)$-$17(12,5) A	&	2.11 	&	202.36 	\\
		345451.103 	&	16(13,3)$-$16(12,4) E	&	1.73 	&	192.34 	\\
		345461.011 	&	28(13,15)$-$27(13,14) E	&	58.68 	&	351.85 	\\
		345465.345 	&	16(13,4)$-$16(12,5) E	&	1.73 	&	192.32 	\\
		345466.962 	&	28(13,16)$-$27(13,15) A	&	58.67 	&	351.86 	\\
		345466.962 	&	28(13,15)$-$27(13,14) A	&	58.67 	&	351.86 	\\
		345486.602 	&	28(13,16)$-$27(13,15) E	&	58.68 	&	351.85 	\\
		345509.021 	&	16(13,4)$-$16(12,5) A	&	1.73 	&	192.34 	\\
		345509.021 	&	16(13,3)$-$16(12,4) A	&	1.73 	&	192.34 	\\
		345532.134 	&	15(13,2)$-$15(12,3) E	&	1.33 	&	182.91 	\\
		345547.616 	&	15(13,3)$-$15(12,4) E	&	1.33 	&	182.89 	\\
		345593.318 	&	15(13,3)$-$15(12,4) A	&	1.33 	&	182.91 	\\
		345593.318 	&	15(13,2)$-$15(12,3) A	&	1.33 	&	182.91 	\\
		345596.828 	&	14(13,1)$-$14(12,2) E	&	0.91 	&	174.07 	\\
		345613.535 	&	14(13,2)$-$14(12,3) E	&	0.91 	&	174.06 	\\
		345647.338 	&	13(13,0)$-$13(12,1) E	&	0.47 	&	165.83 	\\
		345650.835 	&	9(9,1)$-$8(8,1) E	&	3.90 	&	80.31 	\\
		345661.070 	&	14(13,1)$-$14(12,2) A	&	0.91 	&	174.07 	\\
		345661.070 	&	14(13,2)$-$14(12,3) A	&	0.91 	&	174.07 	\\
		345662.771 	&	9(9,0)$-$8(8,0) E	&	3.90 	&	80.33 	\\
		345665.188 	&	13(13,1)$-$13(12,2) E	&	0.47 	&	165.81 	\\
		345714.339 	&	13(13,0)$-$13(12,1) A	&	0.47 	&	165.83 	\\
		345714.339 	&	13(13,1)$-$13(12,2) A	&	0.47 	&	165.83 	\\
		345718.662 	&	9(9,1)$-$8(8,0) A	&	3.90 	&	80.32 	\\
		345718.662 	&	9(9,0)$-$8(8,1) A	&	3.90 	&	80.32 	\\
		345974.664 	&	28(12,16)$-$27(12,15) E	&	61.04 	&	335.43 	\\
		345985.381 	&	28(12,17)$-$27(12,16) A	&	61.05 	&	335.43 	\\
		345985.381 	&	28(12,16)$-$27(12,15) A	&	61.05 	&	335.43 	\\
		346001.616 	&	28(12,17)$-$27(12,16) E	&	61.04 	&	335.43 	\\
		\hline               							
		\multicolumn{4}{c}{\DE\ (v=0)}              									\\
		\cline{1-4}									
		342607.898 	&	19(0,19)$-$18(1,18) EA	&	110.05 	&	167.14 	\\
		342607.898 	&	19(0,19)$-$18(1,18) AE	&	165.08 	&	167.14 	\\
		342607.971 	&	19(0,19)$-$18(1,18) EE	&	440.24 	&	167.14 	\\
		342608.044 	&	19(0,19)$-$18(1,18) AA	&	275.16 	&	167.14 	\\
		343753.320 	&	17(2,16)$-$16(1,15) EA	&	29.01 	&	143.70 	\\
		343753.320 	&	17(2,16)$-$16(1,15) AE	&	58.02 	&	143.70 	\\
		343754.216 	&	17(2,16)$-$16(1,15) EE	&	232.07 	&	143.70 	\\
		343755.112 	&	17(2,16)$-$16(1,15) AA	&	87.02 	&	143.70 	\\
		344357.816 	&	19(1,19)$-$18(0,18) EA	&	55.06 	&	167.18 	\\
		344357.816 	&	19(1,19)$-$18(0,18) AE	&	110.12 	&	167.18 	\\
		344357.929 	&	19(1,19)$-$18(0,18) EE	&	440.51 	&	167.18 	\\
		344358.041 	&	19(1,19)$-$18(0,18) AA	&	165.18 	&	167.18 	\\
		344512.176 	&	11(3,9)$-$10(2,8) EA	&	25.13 	&	72.78 	\\
		344512.219 	&	11(3,9)$-$10(2,8) AE	&	12.57 	&	72.78 	\\
		344515.385 	&	11(3,9)$-$10(2,8) EE	&	100.53 	&	72.78 	\\
		344518.572 	&	11(3,9)$-$10(2,8) AA	&	37.70 	&	72.78 	\\
		\cline{1-4}									
		\multicolumn{4}{c}{\K\ (v=0)}               									\\
		\cline{1-4}									
		343088.615	&	17(5,13)$-$16(5,12) 	&	93.93	&	473.85	\\
		343088.615	&	17(5,12)$-$16(5,11) 	&	93.93	&	473.85	\\
		343172.572	&	17(0,17)$-$16(0,16) 	&	34.28	&	148.30	\\
		343250.411	&	17(4,14)$-$16(4,13) 	&	32.38	&	356.84	\\
		343250.411	&	17(4,13)$-$16(4,12) 	&	32.38	&	356.84	\\
		343376.133	&	17(2,16)$-$16(2,15) 	&	33.81	&	200.53	\\
		343384.676	&	17(3,15)$-$16(3,14) 	&	99.64	&	265.71	\\
		343387.579	&	17(3,14)$-$16(3,13) 	&	99.64	&	265.71	\\
		343693.935	&	17(2,15)$-$16(2,14) 	&	33.81	&	200.61	\\
		\hline
	\end{tabular}
	\begin{tablenotes}
		\item[ ] Notes. The rest frequencies are listed and the transitions are in the form of J(Ka,Kc)$-$J$^{\prime}$(Ka$^{\prime}$,Kc$^{\prime}$). The S$\mu$$^{2}$ is the product of the line strength and the square of the relevant dipole moment. The E$_{\rm up}$ is the upper level energy of each transition. 
	\end{tablenotes}
\end{table}

\section{Molecular Emission}\label{sec:sd}
\begin{figure*}
	\includegraphics[width=\linewidth]{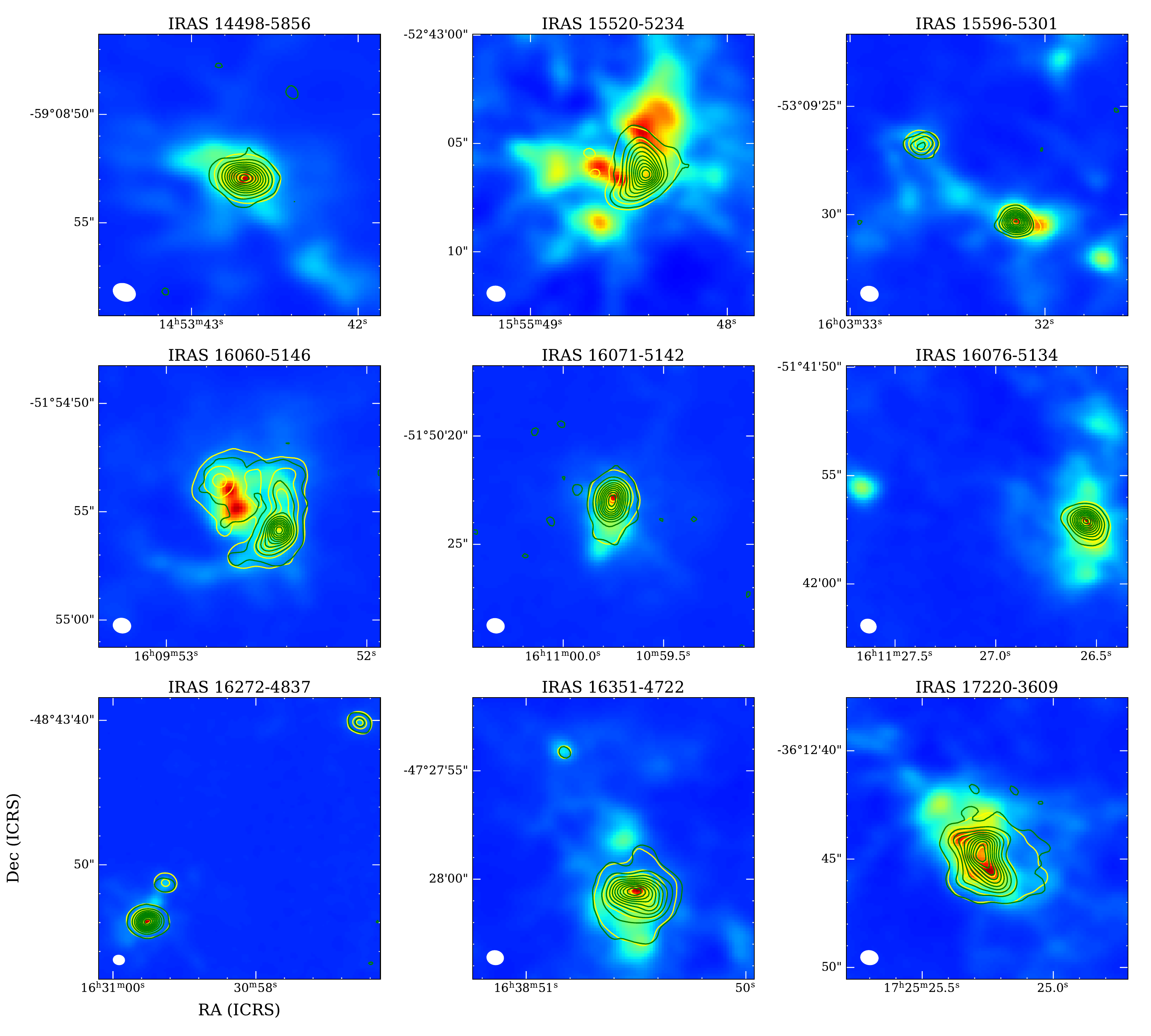}
	\caption{The integrated intensities of \MF\ with different upper energy level transitions. The green and yellow contours represent \MF\ at 342572 MHz (E$_{\rm u}$ = 81 K) and at 342359 MHz (E$_{\rm u}$ = 269 K), respectively. The contour levels are stepped by 10\% of the peak values, with the outermost contour levels as follows: I14498: 5\%, I15520: 3\%, I15596: 8\%, I16060: 2\%, I16071: 2\%, I16076: 6\%, I16272: 2\%, I16351: 2\%, I17220: 2\% of the peak values.} \label{fig9}
\end{figure*}

We compare the molecular emissions of \MF\ with different energy level transitions. The integrated intensity maps of \MF\ at 342572 MHz (E$_{\rm u}$ = 81 K) and at 342359 MHz (E$_{\rm u}$ = 269 K) in 9 high-mass star-forming regions are shown in Figure \ref{fig9}. In the 9 high-mass star-forming regions, there is no difference in the spatial emission between the different energy level transitions of \MF. 

\bsp	
\label{lastpage}
\end{document}